\documentclass[a4paper,11pt]{article}
\pdfoutput=1
\usepackage[T1]{fontenc}
\usepackage{jcappub} 
\usepackage{natbib}
\setcitestyle{square,comma,numbers,sort&compress}
\usepackage{enumerate}
\usepackage{tabularx}
\usepackage[left = 1.0in, top=1.0in,right=1.0in,bottom=1.0in,a4paper]{geometry}
\usepackage[dvipsnames]{xcolor}
\usepackage{comment}
\usepackage[caption=false]{subfig}
\usepackage{cancel}
\usepackage{bbold}

\usepackage[section]{placeins}
\usepackage{listings}
\usepackage{amsbsy}
\usepackage{bm}
\usepackage{tikz}
\usepackage{tikz-feynman}
\usepackage{mathptmx}
\usepackage{pifont}

\tikzfeynmanset{compat=1.0.0}
\usetikzlibrary{shapes.misc}
\tikzset{cross/.style={cross out, draw=black, minimum size=2*(#1-\pgflinewidth), inner sep=0pt, outer sep=0pt},cross/.default={3pt}}
\usetikzlibrary{decorations.pathreplacing, arrows.meta} 
\usepackage{caption}
\usepackage{amsmath}
\newcolumntype{Y}{>{\centering\arraybackslash}X}

\definecolor{myRED}{rgb}{0.8, 0.25, 0.33}
\usepackage[normalem]{ulem}
\usepackage{dsfont}
\usepackage{pbox}
\usepackage{graphicx}
\usepackage{multirow}
\usepackage{tikz}
\usetikzlibrary{arrows}
\usepackage{enumitem} 
\usepackage{ulem}
\usepackage{soul}
\usepackage{lipsum}

\title{\boldmath\huge  
Beyond Neutrino Mass: Observable $n$–$\overline{n}$ Oscillations in UV Complete Seesaw Models
}

\author[a,b]{Ilja Dor\v{s}ner,}
\author[b]{Svjetlana Fajfer,}
\author[b]{Shaikh Saad,}

\emailAdd{dorsner@fesb.hr, svjetlana.fajfer@ijs.si,  shaikh.saad@ijs.si}

\affiliation[a]{University of Split, Faculty of Electrical Engineering, Mechanical Engineering and Naval Architecture, Ru\dj era Bo\v{s}kovi\'{c}a 32, HR-21000 Split, Croatia}

\affiliation[b]{Jožef Stefan Institute, Jamova 39, P.\ O.\ Box 3000, SI-1001 Ljubljana, Slovenia}

\abstract{
Next-generation experiments, such as the Deep Underground Neutrino Experiment and the European Spallation Source, are set to improve sensitivity to neutron-antineutron oscillation, a direct probe of $\Delta B = 2$ baryon number violation, with particularly significant gains expected at the latter. The discovery of such a rare $\Delta B = 2$ process would indicate physics beyond the Standard Model and could point to specific unified theories that allow observable $n-\overline{n}$ transitions. We accordingly examine $n-\overline{n}$ oscillations within a unified framework that accounts for charged fermion masses and generates viable neutrino masses via the seesaw mechanism. More specifically, we show that $n-\overline{n}$ oscillations can arise from two specific topologies within two distinct $SU(5)$ scenarios. One topology requires a presence of two color-sextet scalars in the Type II seesaw framework, whereas the other involves a scalar sextet and a color-octet fermion in the Type III seesaw framework. While the former topology can be realized in the $SO(10)$/Pati-Salam frameworks, the latter finds a natural embedding in $SU(5)$, which constitutes one of the key novelties of our work.  Remarkably enough, the same dynamics responsible for fermion masses also induces baryon number violation, thus linking $n-\overline{n}$ oscillations to the flavor structure of the theory. We show that, given a TeV-scale mass for one of the colored states, upcoming searches for such $\Delta B = 2$ processes can probe for a presence of the other colored states with masses up to $10^{11}\,\mathrm{GeV}$, well beyond the reach of colliders. This positions $n-\overline{n}$ oscillations as a rare low-energy portal to grand unification and ultra-heavy new physics. 
}

\makeatletter
\gdef\@fpheader{}
\makeatother
\begin{document}
\maketitle
\flushbottom

\section{Introduction}
The observation of baryon number $(B)$ violation would provide unambiguous evidence of physics beyond the Standard Model (SM).   Among the most intriguing consequences of grand unified theories (GUTs)~\cite{Pati:1973rp,Pati:1974yy, Georgi:1974sy, Georgi:1974yf, Georgi:1974my, Fritzsch:1974nn} are baryon-number-violating processes. While the hallmark prediction of GUTs is $\Delta B = 1$ proton decay, for example $p \to e^+ \pi^0$, $\Delta B = 2$ processes, such as neutron--antineutron ($n-\overline{n}$) oscillations, may dominate in certain models or specific regions of the parameter space.  Although no such $n-\overline{n}$ transitions have been observed to date, stringent experimental bounds already exist. In particular, Super-Kamiokande (SK) sets a strong lower limit on the $n-\overline{n}$ oscillation time given by $\tau_{n-\overline{n}}^\mathrm{SK}  \geq 4.7\times 10^8\,\mathrm{s}$~\cite{Super-Kamiokande:2020bov}. On the other hand, ILL experiment~\cite{Baldo-Ceolin:1994hzw} provides $\tau_{n-\overline{n}}^\mathrm{ILL}  \geq 8.6\times 10^7\,\mathrm{s}$ from free $n-\overline{n}$ transitions. Future experiments are expected to significantly enhance this sensitivity. For example, the Deep Underground Neutrino Experiment (DUNE)~\cite{DUNE:2015lol} is expected to probe lifetimes exceeding $\tau_{n-\overline{n}}^\mathrm{DUNE}\sim 7 \times 10^8$\,s, while the proposed NNBAR experiment~\cite{Addazi:2020nlz} at the European Spallation Source (ESS) aims to reach an impressive $\tau_{n-\overline{n}}^\mathrm{NNBAR}\sim 3 \times 10^9$\,s. Such unprecedented sensitivity enables the exploration of baryon number violating physics at energy scales far beyond the reach of current or foreseeable collider experiments, firmly establishing $n-\overline{n}$ oscillation searches as a uniquely powerful probe of high-scale new physics.

\begin{figure}[h]
\centering
\begin{minipage}[b]{0.48\textwidth}
\centering
\begin{tikzpicture}
  \begin{scope}[rotate=180]
    \begin{feynman}
      \vertex (a) at (0,0);
      \vertex (x1) at (-1.5, 0); 
      \vertex (x2) at (1.5, 0);   
      \vertex (x3) at (0,-1.5);   
      \vertex (f1) at (-3, 1.6);
      \vertex (f2) at (-3, -1.6);
      \vertex (f3) at (3, 1.6);
      \vertex (f4) at (3, -1.6);

      \vertex (f5) at (-1.5, -3);
      \vertex (f6) at (1.5, -3);

      \diagram* {
        (a) -- [charged scalar, thick,  edge label=\(S_2\)] (x1),
        (a) -- [charged scalar, thick,  edge label'=\(S_1\)] (x2),
        (a) -- [charged scalar, thick,  edge label=\(S_3\)] (x3),

        (f1) -- [fermion, thick, edge label=\(u^c\)] (x1),
        (f2) -- [fermion, thick, edge label'=\(d^c\)] (x1),

        (f3) -- [fermion, thick, edge label=\(u^c\)] (x2),
        (f4) -- [fermion, thick, edge label'=\(d^c\)] (x2),

        (f5) -- [fermion, thick, edge label'=\(d^c\)] (x3),
        (f6) -- [fermion, thick, edge label=\(d^c\)] (x3),
      };
    \end{feynman}
  \end{scope}
\end{tikzpicture}
\vspace{0.2cm}
\caption*{(a) Topology-A}
\end{minipage}
\hfill
\begin{minipage}[b]{0.48\textwidth}
\centering
\begin{tikzpicture}
\begin{feynman}
\vertex (a1) at (-0.8,2.5) {\(d^c\)};
\vertex (a2) at (-3.2,2.5) {\(u^c\)};
\vertex (b1) at (0.9,2.5) {\(u^c\)};
\vertex (b2) at (3.2,2.5) {\(d^c\)};
\vertex (a3) at (-3,-1) {\(d^c\)};
\vertex (b3) at (3,-1) {\(d^c\)};
\coordinate (v1) at (-2,1.5);
\coordinate (v2) at (-1.5,0);
\coordinate (v3) at (2,1.5);
\coordinate (v4) at (1.5,0);

\diagram* {
  (a1) -- [fermion, thick] (v1), 
  (v2) -- [charged scalar, thick, edge label={\(S_1\)}] (v1),
  (a2) -- [fermion, thick] (v1),
  (a3) -- [fermion, thick] (v2),
  (b1) -- [fermion, thick] (v3),
  (v4) -- [charged scalar, thick, edge label'={\(S_2\)}] (v3),
  (b2) -- [fermion, thick] (v3),
  (b3) -- [fermion, thick] (v4),
  (v2) -- [plain, thick, edge label] (v4),
};
\node at (0.1,-0.5) {\(F_1\)};
\end{feynman}
\end{tikzpicture}
\vspace{0.2cm}
\caption*{(b) Topology-B}
\end{minipage}

\caption{  Topology A requires beyond the SM scalars, such as at least two distinct color sextets. Topology B requires both beyond the SM scalars and fermions, for example, at least one color sextet and a colored fermion. } \label{fig:topology}
\end{figure}

Motivated by these remarkable developments and the growing experimental reach, in this work, we investigate two distinct topologies --- Topology A and Topology B --- which emerge naturally within unified theories, as shown in Fig.~\ref{fig:topology}. In particular, we consider two simple models based on $SU(5)$ gauge group that generate viable masses for the SM charged fermions and neutrinos and simultaneously give rise to observable $n-\overline{n}$  oscillation signals. We demonstrate that the topologies shown in Fig.\ref{fig:topology} arise naturally by embedding the seesaw mechanism \cite{Minkowski:1977sc,Yanagida:1979as,Glashow:1979nm,Gell-Mann:1979vob,Mohapatra:1979ia,Schechter:1980gr,Schechter:1981cv} for the neutrino mass generation within an $SU(5)$ grand unified framework. In this work, we consider two concrete realizations of these topologies, namely Model A and Model B, where Model A realizes Type II seesaw within the $SU(5)$ unified framework~\cite{Ma:1998dn,Dorsner:2005fq,Dorsner:2005ii,Dorsner:2006hw,Dorsner:2007fy,Antusch:2022afk,Calibbi:2022wko,Antusch:2023mqe,Kaladharan:2024bop,Dev:2025sox}, while Model B is based on a Type III seesaw embedding~\cite{Ma:1998dn,Bajc:2006ia,Dorsner:2006fx,FileviezPerez:2007bcw,Antusch:2023kli}.  Even though both Type II and Type III seesaws\footnote{ Within the $SU(5)$ framework, the type I seesaw mechanism is less appealing, as it requires introduction of gauge-singlet states.  } have been studied previously in the context of $SU(5)$ GUT, $n-\overline{n}$ oscillations within this context have not been explored, and the parameter space relevant for $n-\overline{n}$ oscillations has not been singled out. Our goal in this work is to rectify this omission. As we will show, when a Type II seesaw is embedded in a $SU(5)$ GUT, Topology A~\footnote{Previously, Topology A of $n-\overline{n}$ oscillation has been studied primarily in the context of Pati–Salam and $SO(10)$ frameworks, or within SM extensions involving scalar diquarks~\cite{Mohapatra:1980qe,Mohapatra:1982xz,Chang:1984qr,Babu:2006xc,Babu:2008rq,Baldes:2011mh,Babu:2012vc,Babu:2012qia,Arnold:2012sd,Bernal:2012gv,Babu:2013yca,Patra:2014goa,Herrmann:2014fha,Saad:2017pqj,Asaka:2019ocw,Girmohanta:2020qfd,Fridell:2021gag,Patel:2022nek}.}, which utilizes two distinct scalar sextets, naturally emerges. In contrast, a realization of a Type III seesaw gives rise to Topology B~\footnote{Topology B has previously been explored in the context of SM extensions involving color-triplet scalars and right-handed neutrinos, as well as within certain supersymmetric frameworks~\cite{Zwirner:1983dgv,Barbieri:1985ty,Mohapatra:1986bd,Lazarides:1986jt,Goity:1994dq,Babu:2001qr,Babu:2006wz,Allahverdi:2010im,Gu:2011ff,Gu:2011fp,Allahverdi:2013mza,Dev:2015uca,Dhuria:2015swa,Ghalsasi:2015mxa,Gu:2016ghu,Calibbi:2016ukt,Gu:2017cgp,Calibbi:2017rab,Allahverdi:2017edd,Grojean:2018fus}.}, involving a scalar sextet and a fermionic color octet. This particular possibility, to our knowledge, has not been considered in the literature before. Notably, the former scenario can be realized within the $SO(10)$/Pati–Salam framework, while the latter finds a natural embedding in the $SU(5)$ setup, representing one of the key novelties of our work.

The scenarios we consider are realistic since they successfully generate viable fermion masses, including neutrinos, achieve consistent gauge coupling unification, and satisfy current proton decay bounds, as well as present LHC and flavor physics constraints.  Remarkably enough, the same dynamics responsible for the fermion mass generation also induces baryon-number violation, thereby linking $n-\overline{n}$ oscillations directly to the flavor structure of the theory.  We demonstrate that, given a TeV-scale mass for one of the colored states, upcoming searches for such $\Delta B = 2$ processes can probe the other colored states with masses up to $10^{11}\,\mathrm{GeV}$, well beyond the reach of current or near-future colliders.   This positions $n-\overline{n}$ oscillations as a unique low-energy window into grand unification and ultra-heavy new physics. We hope that the analysis presented in this work further motivates and inspires future experimental searches for $n-\overline{n}$ oscillations.

Note that, in addition to the rare $\Delta B = 2$ processes, the superheavy gauge bosons as well as some of the color-triplet scalars also induce the usual $\Delta B = 1$ proton decays. The former leads to the most prominent proton decay mode, $p \to \pi^0 e^+$, whereas scalar leptoquarks are expected to dominate the $p \to K^+ \overline{\nu}$ channel. Although these decay rates are not uniquely fixed within these scenarios, both processes could lie within the observable range and could offer a complementary probe for testing these models.

The manuscript is organized as follows. In Sec.~\ref{sec:01}, we present the general model setup. Sec.~\ref{sec:02} contains the construction and details of Model A, while Sec.~\ref{sec:03} provides the corresponding discussion for Model B. We conclude in Sec.~\ref{sec:04}.

\section{Model setup}\label{sec:01}

We analyze potentially observable $n-\overline{n}$ oscillation signatures within two specific $SU(5)$ models that implement the same mechanism for the charged fermion mass generation. We accordingly start by introducing this common feature in the next section.

\subsection{Charged fermion masses}

The Standard Model (SM) fermions comprise three distinct copies of  5-dimensional and 10-dimensional $SU(5)$ representations. The exact embedding of these fermions reads
\begin{align}
&{\overline{5}_F}_i=  d_i^c (\overline{3},1,1/3) + \ell_i (1,2,-1/2) ,  \label{eq:GG1}\\
&{10_F}_i= q_i (3,2,1/6) + u_i^c (\overline{3},1,-2/3) + e_i^c (1,1,1), \label{eq:GG2} 
\end{align} 
where the numbers in the brackets represent transformation properties under the SM gauge group $SU(3) \times SU(2) \times U(1)$ and $i(=1,2,3)$ is the family index. The components of $\overline{5}_F$ and $10_F$, in the $SU(5)$ space, are $\overline{5}_{F\alpha}$ and $10_F^{\alpha \beta}=-10_F^{\beta \alpha}$, where $\alpha,\beta(=1,\ldots,5)$ are $SU(5)$ indices. We employ the usual notation, where, for example, $\ell_1$ in Eq.\ \eqref{eq:GG1} is the $SU(2)$ leptonic doublet of the SM that comprises a charged lepton of the first generation and the associated neutrino, both in the flavor basis. We also introduce subscripts $F$ and $H$ to denote representations containing fermions and scalars, respectively, for clarity of exposition.  

The symmetry breaking of $SU(5)$ down to the SM gauge group is accomplished through a vacuum expectation value (VEV) of a single 24-dimensional scalar representation, where the $SU(5)$ components of that representation are $24_{H \beta}^\alpha$ with $24_{H \alpha}^\alpha=0$, where a summation over repeated indices is understood. The SM decomposition of $24_H$ is
\begin{align}
24_H&=\Phi_1(1,1,0)+\Phi_2(1,3,0)+\Phi_3(8,1,0)+
 \Phi_4(3,2,-5/6)+\Phi_4^*(\overline{3},2,5/6),
\end{align} 
whereas the phenomenologically viable VEV of $24_H$ is defined via 
\begin{align}
\langle 24_{H} \rangle & = v_{24}\; \mathrm{diag}(-1,-1,-1,3/2,3/2). \label{eq:24vev}
\end{align}
This VEV yields the masses for the gauge fields that are associated with the generators that are broken
in the transition from $SU(5)$ to $SU(3) \times SU(2) \times U(1)$ gauge group. These vector bosons are usually denoted with $X$ and $Y$, as they comprise the SM multiplet that transforms as $(3,2,-5/6)$, and their masses are 
\begin{align}
M_{X,Y}=\sqrt{25/8}\; g_\mathrm{GUT}\; v_{24}.
\end{align}
Here, $g_\mathrm{GUT}$ is a gauge coupling constant of $SU(5)$ at the scale of gauge coupling unification $M_\mathrm{GUT} \equiv M_{X,Y}$, where the three gauge couplings of the SM meet.

The scalar sector that accomplishes the breaking of the $SU(3) \times SU(2) \times U(1)$ symmetry down to $SU(3) \times U(1)_\mathrm{em}$ and, consequentially, generates viable masses of the SM charged fermions and gauge bosons comprises a single 5-dimensional and a single 45-dimensional representation\footnote{   An alternative option to correct the wrong mass relations is to add a pair of $5+\overline{5}$~\cite{Babu:2012pb} or $15+\overline{15}$~\cite{Oshimo:2009ia,Dorsner:2019vgf} vectorlike fermions. However, as will be shown later, the presence of a $45_H$ in our setup is essential not only to achieve gauge coupling unification but to also generate observable $n-\overline{n}$ oscillations.   }, where the properties of $45_H$, in the $SU(5)$ space, are $45^{\alpha \beta}_{H \gamma}=-45^{ \beta \alpha}_{H \gamma}$ and $45^{\alpha \beta}_{H \alpha}=0$. 

The decompositions of $5_H$ and $45_H$ into the SM multiplets are  
\begin{align}
5_H&=\phi_2(3,1,-1/3)+\phi_1(1,2,1/2),
\\
45_H& =\Sigma_1 (1,2,1/2) + \Sigma_2 (3,1,-1/3) + \Sigma_3 (\overline{3},1,4/3)
+ \Sigma_4 (\overline{3},2,-7/6) + \Sigma_5 (3,3-1/3) 
\nonumber \\ & 
 ~~~+ \Sigma_6 (\overline{6},1,-1/3) + \Sigma_7 (8,2,1/2),  
\end{align}
whereas the VEVs can be succinctly defined via
\begin{align} \langle \phi^5 \rangle  = v_{5}/\sqrt{2},\qquad 
\langle\Sigma^{45}_4\rangle = \sqrt{3}\; v_{45}/4,\;   \langle\Sigma^{15}_1\rangle = \langle\Sigma^{25}_2\rangle=\langle\Sigma^{35}_3\rangle ,
\end{align}
such that $|v_5|^2+|v_{45}|^2=(246\,\mathrm{GeV})^2$.  Without loss of generality, one can always take $v_5$ to be real. On the other hand, $v_{45}$ can still be complex. However, its phase is irrelevant for the fermion mass fit, as it can always be absorbed into the corresponding Yukawa coupling matrices. Therefore, for simplicity, we take $v_{45}$ also to be real.  

The Yukawa sector relevant for the charged fermion mass generation reads
\begin{equation*}
\mathcal{L}_{Y} = Y_{1ij} \overline{5}_{Fi\alpha}  10_{Fj}^{\alpha \beta} 
  5^*_{H\beta}  +  Y_{2ij}
 \overline{5}_{Fi\delta} 10_{Fj}^{\alpha \beta} 45^{*\delta}_{H\alpha \beta}  
+\epsilon_{\alpha \beta \gamma \delta \eta} \left( Y_{3ij} 
10_{Fi}^{\alpha \beta}  10_{Fj}^{\gamma \delta}  5_{H}^\eta 
+ Y_{4ij}  10_{Fi}^{\alpha \beta} 10_{Fj}^{\zeta \gamma}
45_{H\zeta}^{\delta \eta} \right),
\end{equation*}
where we explicitly specify contractions in both the $SU(5)$ and flavor spaces ($\varepsilon$ here is the five-index Levi-Civita tensor).
These contractions, in turn, yield mass matrices at the GUT scale for the charged leptons, down-type quarks, and up-type quarks that are
\begin{eqnarray}
M_E &=& \frac{1}{2} Y_1  v_5 + \frac{\sqrt{3}}{2\sqrt{2}}  Y_2  v_{45} ,\label{eq:ME}\\
M_D &=& \frac{1}{2} Y^T_1 v_5 - \frac{1}{2\sqrt{6}}  Y^T_2 v_{45} ,\label{eq:MD}\\
M_U &=& \sqrt{2} \left(Y_3+Y_3^T\right)  v_5  +\frac{1}{\sqrt{3}} \left(Y_4-Y_4^T\right) v_{45} ,\label{eq:MU}
\end{eqnarray}
where, $Y_1$, $Y_2$, $Y_3$, and $Y_4$ are $3 \times 3$ Yukawa coupling matrices with complex entries. 

The mass matrices of the SM fermions are diagonalized via
\begin{align}
&E^T_c M_E E=  M_E^\mathrm{diag}, \label{ME}\\ 
&D^T_c M_D D=  M_D^\mathrm{diag},  \label{MD}\\
&U^T_c M_U U=  M_U^\mathrm{diag}, \label{MU}\\
&N^T M_N N=  M_N^\mathrm{diag}, \label{MN}
\end{align}
where $E_c$, $E$, $D_c$, $D$, $U_c$, $U$, and $N$ are $3 \times 3$ unitary matrices that provide transition from the flavor to the mass eigenstate basis. Note that we also introduce a neutrino mass matrix $M_N$ in Eq.\ \eqref{MN}, where we take neutrinos to be of Majorana nature.

To generate viable $M_N$, we resort to two distinct models that, consequently, produce differing parametric dependence of the $n-\overline{n}$ oscillation signal. The first one uses a 15-dimensional scalar representation to produce neutrino masses via type II seesaw mechanism. We refer to it as Model A in what follows. The second model uses a 24-dimensional fermionic representation to generate neutrino masses via combination of type I and type III seesaw mechanisms. We refer to it as Model B. Note that both models generate charged fermion masses through the VEVs of $5_H$ and $45_H$, where the explicit forms of $M_E$, $M_D$, and $M_U$ are given in Eqs.\ \eqref{eq:ME}, \eqref{eq:MD}, and \eqref{eq:MU}, respectively. 

Even though these two models have been investigated before in great detail, there still does not exist a dedicated study of a particular part of parameter space, if it even exists, where these models could yield  potentially observable $n-\overline{n}$ oscillations. We intend to remedy that in what follows.

\section{Model A}\label{sec:02}

\subsection{Neutrino mass}

\begin{figure}[t!]
\centering
\scalebox{1.2}{
\begin{tikzpicture}[
  line width=0.8pt,
  decoration={markings, mark=at position 0.5 with {\arrow{>}}} 
]
  \draw[black, dashed, postaction={decorate}] (0,2) -- (0,0) node[midway, right] {$\Delta_1\subset 15_H$};
  \draw[black, postaction={decorate}] (-2,0) -- (0,0) node[midway, above] {$\nu\subset \overline 5_F$};
  \draw[black, postaction={decorate}] (2,0) -- (0,0) node[midway, above] {$\nu\subset \overline 5_F$};
  \draw[black, dashed, postaction={decorate}] (-1,3) -- (0,2) node[midway, above left, xshift=-0.5cm] {$\phi_1\subset 5_H$}; 
  \draw[black, dashed, postaction={decorate}] (1,3) -- (0,2) node[midway, above right, xshift=0.5cm] {$\phi_1\subset 5_H$}; 
\node at (-0.2,0) [below right] {\(Y_\nu\)};
\node at (-0.2,2) [above right] {\(\mu\)};
\end{tikzpicture}
}
\caption{Schematic diagram of the neutrino mass generation in Model A.} \label{fig:nuMass-A}
\end{figure}
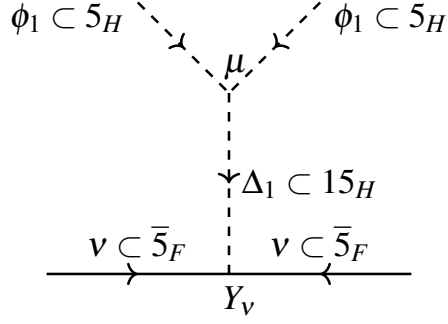

Model A generates neutrino masses  through a presence of a single 15-dimensional scalar representation $15_H^{\alpha \beta} (= 15_H^{ \beta \alpha})$ that decomposes as 
\begin{align}   &15_{H}=\Delta_{1}(1,3,1)+\Delta_{3}(3,2,1/6)+\Delta_{6}(6,1,-2/3).
\end{align}
The only direct interaction between $15_H$ and the SM fermions is
\begin{align}
&\mathcal{L}_Y\supset Y_{\nu ij}\; \overline{5}_{Fi\alpha} \overline{5}_{Fj\beta} 15_H^{\alpha \beta},
\label{eq:typeII}
\end{align}
where $Y_\nu$ is a symmetric matrix in the flavor space. To complete the neutrino mass generation diagram that is depicted in Fig.\ \ref{fig:nuMass-A}, an electrically neutral component of $\Delta_{1}(1,3,1) \in 15_H$ needs to get a VEV through the following contraction 
\begin{align}
V\supset \mu\, 5_H^\alpha 5_H^\beta 15_{H \alpha \beta}^* 
+\mathrm{h.c.}. 
\label{eq:vertex}
\end{align}
The relevant VEV reads 
\begin{align}
\langle 15_H^{55} \rangle = v_{15}&= -\frac{v^2_5}{2m^2_{\Delta_1}} \mu,
\label{eq:VEV_15}
\end{align}
where we observe that $v_{15}$ is restricted by the electroweak precision data (upper bound) and perturbativity of the $Y_\nu$ entries (lower bound) to be within the range: $\mathcal{O}(1)\,\mathrm{GeV}> v_{15} \gtrsim 5\times 10^{-11}\,\mathrm{GeV}$.  Here, the upper bound on this VEV arises from electroweak precision measurements, whereas the lower bound corresponds to the  perturbativity of the Yukawa couplings responsible for generating neutrino masses. Since this VEV is restricted to be subdominant, it does not, for all practical purposes, alter the above relation between $v_{5}$ and $v_{45}$. Finally, Eqs.\ \eqref{eq:typeII} and \eqref{eq:VEV_15} produce the following neutrino mass term
\begin{align}
&\mathcal{L}_Y\supset 2 v_{15}\; \nu_i Y_{\nu ij}  \nu_j =\nu_i M_{Nij} \nu_j,
\label{eq:nu_mass_A}
\end{align}
where, again, $M_N(=2 v_{15} Y_\nu)$ is a neutrino mass matrix in the flavor basis. The neutrino mixing parameters are then determined from the Pontecorvo-Maki-Nakagawa-Sakata (PMNS) mixing matrix defined as $U_\mathrm{PMNS}= E^\dagger N$. 

\subsection{$n-\overline{n}$\, oscillation}

Model A generates $n-\overline{n}$ oscillations via diagram given in Fig.\ \ref{fig:nnBAR-A} that corresponds to Topology A of Fig\ \ref{fig:topology}. As can be seen from this diagram, the $n-\overline{n}$ transition is facilitated by the presence of two scalar sextets. The existence of one sextet, $\Sigma_6$, is required to consistently generate viable charged fermion masses, while the other sextet, $\Delta_6$, is present as a part of the neutrino mass generation mechanism in this framework.

It is important to point out that the model setup discussed in Sec.~\ref{sec:01}, which addresses consistent charged-fermion mass generation but by itself cannot account for neutrino masses, retains $B-L$ as an accidental symmetry of the theory, where $L$ is the lepton number. However, extensions of this basic setup that incorporate Majorana neutrino masses eventually break $B-L$ due to the additional interactions. For example, non-zero Majorana neutrino masses require interaction vertices that violate $\Delta L = 2$, whereas $n-\overline{n}$ oscillations involve interactions with $\Delta B = 2$. Therefore, the process of  $n-\overline{n}$ transition is a consequence of breaking $B-L$ by two units. 
\begin{figure}[th!]
\centering
\begin{tikzpicture}
  \begin{feynman}
    \vertex (a) at (0,0);

    \vertex (x1) at ({1.5*cos(135)}, {1.5*sin(135)});  
    \vertex (x2) at ({1.5*cos(45)}, {1.5*sin(45)});   
    \vertex (x3) at (0,-1.5);                         
    \vertex (f1) at ({3*cos(145 + 16)}, {3*sin(145 + 16)});
    \vertex (f2) at ({3*cos(130 - 18)}, {3*sin(130 - 18)});
    \vertex (f3) at ({3*cos(50 + 20)}, {3*sin(50 + 20)});
    \vertex (f4) at ({3*cos(40 - 20)}, {3*sin(40 - 20)});
    \vertex (f5) at ({-1.5}, -3);
    \vertex (f6) at ({1.5}, -3);

    \diagram* {
      (a) -- [charged scalar, thick,  edge label=\(\Sigma^*_6\)] (x1),
      (a) -- [charged scalar, thick,  edge label'=\(\Sigma^*_6\)] (x2),
      (a) -- [charged scalar, thick,  edge label=\(\Delta_6\)] (x3),

      (f1) -- [fermion, thick, edge label=\(u^c\)] (x1),
      (f2) -- [fermion, thick, edge label'=\(d^c\)] (x1),

      (f3) -- [fermion, thick, edge label=\(u^c\)] (x2),
      (f4) -- [fermion, thick, edge label'=\(d^c\)] (x2),

      (f5) -- [fermion, thick, edge label'=\(d^c\)] (x3),
      (f6) -- [fermion, thick, edge label=\(d^c\)] (x3),
    };
  \end{feynman}
\node at (-0.2,0.8) [below right] {\( \widetilde\mu \)};  
\node at (-0.4,-1.7) [below right] {\( Y^{dd}_{11} \)}; 
\node at (-1.8,1.7) [below right] {\( Y^{ud}_{11} \)};  
\node at (1,1.7) [below right] {\( Y^{ud}_{11} \)};  
\end{tikzpicture}
\caption{ $n-\overline{n}$ oscillation diagram in Model A. } \label{fig:nnBAR-A}
\end{figure}

The bottom vertex in Fig.\ \ref{fig:nnBAR-A} originates from the $SU(5)$ contraction featured in Eq.\ \eqref{eq:typeII} that, at the SM level, yields 
\begin{align}
&\mathcal{L}_Y\supset d^c_i  \left( \sqrt{2} D^T_c Y_\nu D_c \right)_{ij} d^c_j \;\Delta_6=d^c_i (D^T_c M_N D_c)_{ij} d^c_j \;\Delta_6/(\sqrt{2} v_{15})=d^c_i Y^{dd}_{ij} d^c_j \;\Delta_6,
\label{eq:interaction_A}
\end{align}
where $Y^{dd} (=\sqrt{2} D^T_c Y_\nu D_c)$ is written in the mass eigenstate basis and represents the interaction strength between the $d^c$-$d^c$ pairs and a color sextet $\Delta_6$. (Note that  $d^c_i Y_{ij}^{dd} d^c_j \Delta_6$ actually stands for $d^{c T}_i \mathcal{C} Y_{ij}^{dd} d^c_j \Delta_6$, where $\mathcal{C}$ is the usual charge conjugation matrix.) The fact that $Y_\nu$ and, consequently, $M_N$ are directly related to  $Y^{dd}$ represents one of the most relevant constrains on the parameter space that can yield observable $n-\overline{n}$ oscillation signal within the Model A framework.

The trilinear scalar interaction vertex in Fig.\ \ref{fig:nnBAR-A}, on the other hand, is due to the $SU(5)$ contraction
\begin{align}
    V\supset \widetilde \mu\;  45^{*\gamma}_{H \alpha \eta} 45^{*\eta}_{H \delta \gamma} 15^{\alpha \delta}_H \supset \widetilde \mu\; \Sigma^*_6 \Sigma^*_6 \Delta_6,  
\label{eq:vertex2}
\end{align}
where $\widetilde \mu$ is a dimensionful parameter. We assume that $\widetilde \mu \lesssim 3\times \mathrm{max}(m_{\Delta_6}, m_{\Sigma_6})$ to avoid the charge breaking minima. This issue, for example, has been studied in Refs.~\cite{Frere:1983ag,Alvarez-Gaume:1983drc,Casas:1996de} for several specific scenarios.

The remaining two vertices of Fig.\ \ref{fig:nnBAR-A} correspond to an interaction between scalar multiplet  $\Sigma_6 \in 45_H$ and the $u^c$-$d^c$ pairs. More specifically, this interaction reads
\begin{align}\,
    \mathcal{L}_{Y}\supset Y_{2ij}\; 10_{Fi} \,\overline 5_{Fj}\, 45^*_H  \supset  u^c_i \,(U^T_c Y_2 D_c)_{ij} d^c_j \;\Sigma^*_6  = 
     u^c_i\, Y^{ud}_{ij}\, d^c_j \;\Sigma^*_6,
\label{eq:Yud}     
\end{align}
where $Y_2$ and, consequentially, $Y^{ud}(=U^T_c Y_2 D_c)$ are both directly related to an observed mismatch between the masses of the down-type quarks and charged leptons via Eqs.\ \eqref{ME} and \eqref{MD}.

The $d=9$ operator that is relevant for $n-\overline{n}$ oscillations, in the $SU(3) \times U(1)_\mathrm{em}$ invariant basis, is given by~\cite{Fridell:2021gag}
\begin{align}
&\mathcal{O}^2_{RRR}= \left( u^T_iCP_Rd_j \right)  \left( u^T_kCP_Rd_l \right) \left( d^T_mCP_Rd_n \right) T^{SSS}_{ \{ij\}\{kl\}\{mn\} },  
\\&
    T^{SSS}_{ \{ij\}\{kl\}\{mn\} }= \epsilon_{ikm}\epsilon_{jln}+ \epsilon_{jkm}\epsilon_{iln} + \epsilon_{ilm}\epsilon_{jkn} + \epsilon_{jlm}\epsilon_{ikn}.
\end{align}
Following  Ref.~\cite{Fridell:2021gag}, we find the corresponding relevant hadronic matrix element for this operator to be $\langle \overline n |\mathcal{O}^2_{RRR}| n \rangle = -3 \mathcal{M}_1(\mu)/20$, and  take $\mathcal{M}_1(2\,\mathrm{GeV}) = -4.6\times 10^{-4} \,\mathrm{GeV}^6$.

To summarize, the parameters that are featured in Fig.\ \ref{fig:nnBAR-A}, in the physical basis, are
\begin{align}
    \mathcal{L}_Y&\supset Y^{ud}_{ij}\; u^c_i d^c_j\,\Sigma^*_6 + Y^{dd}_{ij}\; d^c_i d^c_j\,\Delta_6,\\
    V &\supset m^2_{\Delta_6} \;\Delta^{dd*}_6\,\Delta^{dd}_6 + m^2_{\Sigma_6} \;\Sigma^*_6 \,\Sigma_6 + \widetilde \mu\;   \Sigma^*_6 \,\Sigma^*_6 \, \Delta_6, 
\end{align}
where it is only $Y^{dd}_{11}$ and $Y^{ud}_{11}$ elements of $Y^{dd}$ and $Y^{ud}$, respectively, that are relevant for $n-\overline{n}$ oscillation signal. All this, when put together, leads to an effective lagrangian of the form
\begin{align}
  \mathcal{L}^{n-\overline{n}}_\mathrm{eff} = - \frac{ Y^{dd}_{11} \left( Y^{ud}_{11} \right)^2 \widetilde \mu }{ m^2_{\Delta_6} \,m^4_{\Sigma_6}}   \mathcal{O}^2_{RRR}. 
\end{align}
Finally, the expression of interest for an inverse of the the $n-\overline{n}$ oscillation time is
\begin{align}
    \tau^{-1}_{n-\overline{n}}= \langle \overline n |\mathcal{L}^{n-\overline{n}}_\mathrm{eff}| n \rangle  = \left| \frac{3\; Y^{dd}_{11} \left( Y^{ud}_{11} \right)^2 \;\widetilde \mu }{20\; m^2_{\Delta_6}\, m^4_{\Sigma_6}}   \mathcal{\widetilde M}_1(2\,\mathrm{GeV})\right|.
\end{align}

We note that the current experimental bound, as given by Super-Kamiokande (SK)~\cite{Super-Kamiokande:2020bov}, is 
\begin{align}
    \tau_{n-\overline{n}}^\mathrm{SK}  \geq 4.7\times 10^8\,\mathrm{s},
\end{align}
whereas the future DUNE~\cite{DUNE:2015lol} and NNBAR~\cite{Addazi:2020nlz} experiments are expected to yield 
\begin{align}
    \tau_{n-\overline{n}}^\mathrm{DUNE}  \geq 7\times 10^8\,\mathrm{s}, \quad \;\;\; \tau_{n-\overline{n}}^\mathrm{NNBAR}  \geq 3\times 10^9\,\mathrm{s}.
\end{align}
If we take into consideration the most stringent current bound and future expectations, we find that  
\begin{align}
\mathrm{SK:}\;\;\;   &\frac{M}{\mathrm{1\,TeV}} \geq  5.5\times 10^2 \left[  |Y^{dd}_{11}| | Y^{ud}_{11} |^2 \right]^{1/5},  
\\
\mathrm{DUNE:}\;\;\;   &\frac{M}{\mathrm{1\,TeV}} \geq  6.0\times 10^2 \left[ |Y^{dd}_{11}| | Y^{ud}_{11} |^2 \right]^{1/5},  
\\
\mathrm{NNBAR:}\;\;\;   &\frac{M}{\mathrm{1\,TeV}} \geq  8.0\times 10^2 \left[ |Y^{dd}_{11}| | Y^{ud}_{11} |^2 \right]^{1/5},  
\end{align}
where we take  $\widetilde \mu= m_{\Delta_6}$ and set, for simplicity, $m_{\Delta_6}= m_{\Sigma_6} \equiv M$. 

The complete Model A parametrization of $\tau_{n-\overline{n}}$ yields the following observable window of the $n-\overline{n}$ oscillation signal in near future
\begin{align}
&
3.146\times 10^{14} \; \mathrm{(NNBAR)} \geq 
\frac{1}{|Y^{dd}_{11}| | Y^{ud}_{11} |^2  }  
  \left( \frac{m_{\Delta_6}}{1\,\mathrm{TeV}}  \right)^2
  \left( \frac{m_{\Sigma_6}}{1\,\mathrm{TeV}}  \right)^4
 \left(   \frac{1\,\mathrm{TeV}}{\widetilde \mu}  \right)  
   \geq    4.93\times 10^{13}\; \mathrm{(SK)}. \label{eq:fitnnBAR:A}  
\end{align}
The lower bound in Eq.\ \eqref{eq:fitnnBAR:A}, as indicated, is given by the current SK limit~\cite{Super-Kamiokande:2020bov}, whereas the upper bound corresponds to the NNBAR expected sensitivity. 

What we need to address next is whether Model A can indeed yield this window of observability, i.e., $4.7\times 10^8\,\mathrm{s}   \leq  \tau_{n-\overline{n}}  \leq 3\times 10^9\,\mathrm{s}$.  

\subsection{Numerical analysis}

We aim to find a viable parameter space for observable $\tau_{n-\overline{n}}$ within the $m_{\Delta_6}$-$m_{\Sigma_6}$ plane for those values of $Y^{dd}_{11}$ and $Y^{ud}_{11}$ that are in agreement with all potentially relevant flavor physics constraints as well as the SM fermion mass generation within the Model A framework. In order to do that we first discuss these constraints and then perform numerical fit of the SM fermion masses to obtain corresponding values of $Y^{dd}_{11}$ and $Y^{ud}_{11}$. We note that entries in $Y^{dd}$ and $Y^{ud}$ scale with $v_{15}$ and $v_{45}$, respectively. We thus require, in order to perform the most conservative parameter space analysis, that the largest absolute values of entries in matrices $Y_1$, $Y_2$, $Y_3$, and $Y_\nu$ do not exceed one. If the entries of either $Y^{dd}$ or $Y^{ud}$ are scaled down with respect to the values that we intend to consider, the observable $n-\overline{n}$ oscillation signal would accordingly probe even lower values of masses of $\Delta_6$ and $\Sigma_6$ scalars.

\subsubsection{Constraints from gauge coupling unification}
We first observe that the phenomenologically viable gauge coupling unification requires $\Sigma_5(3,3,-1/3)\in 45_H$ to be several orders of magnitude below $10^{12}$\,GeV due to the fact that both $m_{\Delta_6}$ and $m_{\Sigma_6}$ need to be significantly below the gauge coupling unification scale. But, if $m_{\Sigma_5}$ is indeed below $10^{12}$\,GeV, $\Sigma_5$ would mediate proton decay unless one forbids or sufficiently suppresses its leptoquark couplings due to the $10_F$-$10_F$-$45_H$ contraction. We accordingly assume that $Y_4$ of Eq.\ \eqref{MU} is a null-matrix. This, in turn, means that $M_U$ is a symmetric matrix and, consequently, that $U_c$ can be written as 
\begin{align}
U_c=U^*=  D^*  Q  V_\mathrm{CKM}^T P,
\label{eq:U_c}
\end{align}
where we use $U^\dagger D = P V_\mathrm{CKM} Q$ with $P= \mathrm{diag}(e^{i \omega_1},e^{i \omega_2},e^{i \omega_3})$ and $Q= \mathrm{diag}(e^{i \omega_4},e^{i \omega_5},1)$. Here, $V_\mathrm{CKM}$ is the Cabibbo-Kobayashi-Maskawa (CKM) matrix of the SM with experimentally known entries.

\subsubsection{Constraints from collider data and flavor physics}

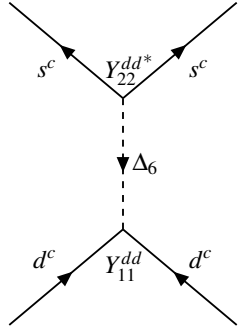
\begin{figure}[h]
\centering
\resizebox{0.2\textwidth}{!}{
\begin{tikzpicture}
  \begin{feynman}
    \vertex (a) at (0,-1);      
    \vertex (b) at (0,1);        
    \vertex (f1) at ({-3*sin(35)}, {-3*cos(35)});  
    \vertex (f2) at ({3*sin(35)}, {-3*cos(35)});  
    \vertex (f3) at ({-3*sin(35)}, {3*cos(35)});   
    \vertex (f4) at ({3*sin(35)}, {3*cos(35)});    

    \diagram* {
      (b) -- [charged scalar, thick, edge label=\(\Delta_6\)] (a),

      (f1) -- [fermion, thick, edge label=\(d^c\)] (a),
      (f2) -- [fermion, thick, edge label'=\(d^c\)] (a),

      (b) -- [fermion, thick, edge label=\(s^c\)] (f3),
      (b) -- [fermion, thick, edge label'=\(s^c\)] (f4),
    };
  \end{feynman}
\node at (-0.4,-1.2) [below right] {\( Y^{dd}_{11} \)}; 
\node at (-0.4,1.8) [below right] {\( {Y^{dd}_{22}}^* \)}; 
\end{tikzpicture} 
}
\caption{Tree-level $K^0-\overline K^0$ oscillation present in Model A. Analogous diagrams yield  $B_{d,s}^0-\overline B_{d,s}^0$ oscillations.} \label{mmBAR-tree}
\end{figure}

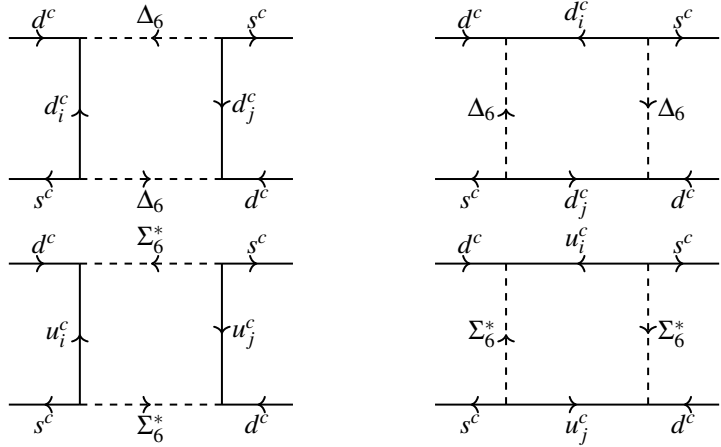
\begin{figure}[ht]
\centering
\resizebox{0.6\textwidth}{!}{
\begin{tikzpicture}[
  line width=0.8pt,
  decoration={
    markings,
    mark=at position 0.5 with {\arrow{>}}
  },
]
  \draw[black, postaction={decorate}] (-6,1) -- (-5,1) node[midway, above] {$d^c$};
  \draw[black, postaction={decorate}] (-5,-1) -- (-6,-1) node[midway, below] {$s^c$};
  \draw[black, postaction={decorate}] (-3,1) -- (-2,1) node[midway, above] {$s^c$};
  \draw[black, postaction={decorate}] (-2,-1) -- (-3,-1) node[midway, below] {$d^c$};
  \draw[black, postaction={decorate}] (-5,-1) -- (-5,1) node[midway, left] {$d_i^c$};
  \draw[black, postaction={decorate}] (-3,1) -- (-3,-1) node[midway, right] {$d_j^c$};
  \draw[black, dashed, postaction={decorate}] (-3,1) -- (-5,1) node[midway, above] {$\Delta_6$};
  \draw[black, dashed, postaction={decorate}] (-5,-1) -- (-3,-1) node[midway, below] {$\Delta_6$};

  \draw[black, postaction={decorate}] (0,1) -- (1,1) node[midway, above] {$d^c$};
  \draw[black, postaction={decorate}] (1,-1) -- (0,-1) node[midway, below] {$s^c$};
  \draw[black, postaction={decorate}] (3,1) -- (4,1) node[midway, above] {$s^c$};
  \draw[black, postaction={decorate}] (4,-1) -- (3,-1) node[midway, below] {$d^c$};
  \draw[black, postaction={decorate}] (3,1) -- (1,1) node[midway, above] {$d_i^c$};
  \draw[black, postaction={decorate}] (1,-1) -- (3,-1) node[midway, below] {$d_j^c$};
  \draw[black, dashed, postaction={decorate}] (1,-1) -- (1,1) node[midway, left] {$\Delta_6$};
  \draw[black, dashed, postaction={decorate}] (3,1) -- (3,-1) node[midway, right] {$\Delta_6$};

  \draw[black, postaction={decorate}] (-6,-2.2) -- (-5,-2.2) node[midway, above] {$d^c$};
  \draw[black, postaction={decorate}] (-5,-4.2) -- (-6,-4.2) node[midway, below] {$s^c$};
  \draw[black, postaction={decorate}] (-3,-2.2) -- (-2,-2.2) node[midway, above] {$s^c$};
  \draw[black, postaction={decorate}] (-2,-4.2) -- (-3,-4.2) node[midway, below] {$d^c$};
  \draw[black, postaction={decorate}] (-5,-4.2) -- (-5,-2.2) node[midway, left] {$u_i^c$};
  \draw[black, postaction={decorate}] (-3,-2.2) -- (-3,-4.2) node[midway, right] {$u_j^c$};
  \draw[black, dashed, postaction={decorate}] (-3,-2.2) -- (-5,-2.2) node[midway, above] {$\Sigma_6^*$};
  \draw[black, dashed, postaction={decorate}] (-5,-4.2) -- (-3,-4.2) node[midway, below] {$\Sigma_6^*$};

  \draw[black, postaction={decorate}] (0,-2.2) -- (1,-2.2) node[midway, above] {$d^c$};
  \draw[black, postaction={decorate}] (1,-4.2) -- (0,-4.2) node[midway, below] {$s^c$};
  \draw[black, postaction={decorate}] (3,-2.2) -- (4,-2.2) node[midway, above] {$s^c$};
  \draw[black, postaction={decorate}] (4,-4.2) -- (3,-4.2) node[midway, below] {$d^c$};
  \draw[black, postaction={decorate}] (3,-2.2) -- (1,-2.2) node[midway, above] {$u_i^c$};
  \draw[black, postaction={decorate}] (1,-4.2) -- (3,-4.2) node[midway, below] {$u_j^c$};
  \draw[black, dashed, postaction={decorate}] (1,-4.2) -- (1,-2.2) node[midway, left] {$\Sigma_6^*$};
  \draw[black, dashed, postaction={decorate}] (3,-2.2) -- (3,-4.2) node[midway, right] {$\Sigma_6^*$};
\end{tikzpicture}
}
\caption{One-loop $K^0-\overline K^0$ oscillation due to either $\Delta_6$ or $\Sigma_6$ exchange. Analogous diagrams generate  $B_{d,s}^0-\overline B_{d,s}^0$ and $D^0-\overline D^0$ oscillations.} \label{mmBAR-loop}
\end{figure}

We note that the LHC mass limits on scalars that couple to quark-quark pairs are rather stringent and can go up to $7.5$\,TeV \cite{CMS:2019gwf} depending on the associated branching ratios. Interestingly, future projections for such diquark resonances at the High-Luminosity LHC (HL-LHC) and the Future Circular Collider (FCC-hh) extend the discovery prospect up to $9.4$\,TeV and $63$\,TeV for integrated luminosities of $3$\,ab$^{-1}$ and $30$\,ab$^{-1}$~\cite{Harris:2022kls,Bernardi:2022hny}, respectively.   We take these limits into account when we perform our numerical fit and check for the Model A viability. Also, in Model A we have  scalar color sextets $\Delta_6$ and $\Sigma_6$ that couple to $d^c$-$d^c$ and $u^c$-$d^c$ pairs, respectively. This, consequentially, can lead to meson-antimeson oscillations through diagrams given in Figs.\ \ref{mmBAR-tree} and \ref{mmBAR-loop}. For the $\Delta_6$ sextet, tree-level processes provide the most stringent constraints on the parameter space, while box diagrams put bounds on different combinations of Yukawa couplings. In contrast, for the $\Sigma_6$ sextet, no such tree-level processes are relevant and box diagrams provide the leading bounds.  We implement constraints from these processes in our numerical analysis. The limits in question are taken from Ref.\ \cite{Fortes:2013dba} (for model independent bounds, see, for example, Ref.~\cite{Isidori:2023sma}) and summarized in Table~\ref{tab:constraints}.

\begin{table}[t!]
\centering
\resizebox{0.55\textwidth}{!}{
\begin{tabular}{|c|c|c|}
\hline
Process & Diagram & Constraint \\
\hline\hline

& Tree & $\left|Y^{dd}_{22}{Y^{dd}_{33}}^*\right| \leq 7.04 \times 10^{-4} \left(m_{\Delta_{6}}/1\,\mathrm{ TeV}\right)^2$ 
\\
$\Delta m_{B_s}$ & Box & $\sum_{i=1}^3 \left|Y^{dd}_{i3}{Y^{dd}_{i2}}^*\right| \leq 0.14 \left(m_{\Delta_{6}}/1\,\mathrm{ TeV}\right)$ 
\\
& Box & $\sum_{i=1}^3 \left|Y^{ud}_{i3}{Y^{ud}_{i2}}^*\right| \leq 1.09 \left(m_{\Sigma_{6}}/1\,\mathrm{ TeV}\right)$ 
\\
\hline\hline
& Tree & $\left|Y^{dd}_{11}{Y^{dd}_{33}}^*\right| \leq 2.75 \times 10^{-5} \left(m_{\Delta_{6}}/1\,\mathrm{ TeV}\right)^2$ \\
$\Delta m_{B_d}$ & Box & $\sum_{i=1}^3 \left|Y^{dd}_{i3}{Y^{dd}_{i1}}^*\right| \leq 0.03 \left(m_{\Delta_{6}}/1\,\mathrm{ TeV}\right)$ 
\\
& Box & $\sum_{i=1}^3 \left|Y^{ud}_{i3}{Y^{ud}_{i1}}^*\right| \leq 0.21 \left(m_{\Sigma_{6}}/1\,\mathrm{ TeV}\right)$ 
\\
\hline\hline
& Tree & $\left|Y^{dd}_{11}{Y^{dd}_{22}}^*\right| \leq 6.56 \times 10^{-6} \left(m_{\Delta_{6}}/1\,\mathrm{ TeV}\right)^2$ \\
$\Delta m_{K}$ & Box & $\sum_{i=1}^3 \left|Y^{dd}_{i2}{Y^{dd}_{i1}}^*\right| \leq 0.01 \left(m_{\Delta_{6}}/1\,\mathrm{ TeV}\right)$ 
\\
& Box & $\sum_{i=1}^3 \left|Y^{ud}_{i1}{Y^{ud}_{i2}}^*\right| \leq 0.10 \left(m_{\Sigma_{6}}/1\,\mathrm{ TeV}\right)$ 
\\ \hline

\end{tabular}
}
\caption{Relevant flavor physics bounds inferred from Ref.\ \cite{Fortes:2013dba} (for model independent bounds, see, for example, Ref.~\cite{Isidori:2023sma}). For the $\Delta_6$ sextet, tree-level processes provide the most stringent constraints on the parameter space, while box diagrams put bounds on different combinations of Yukawa couplings. In contrast, for the $\Sigma_6$ sextet, no such tree-level processes are relevant and box diagrams provide the leading constraints.   }
\label{tab:constraints}
\end{table}

\subsubsection{Numerical Fit}

Fermion mass fit, for Model A, is performed at the gauge coupling unification scale $M_\mathrm{GUT}$, with the input values at $M_\mathrm{GUT}$ that are taken from Ref.\ \cite{Babu:2016bmy} for convenience. We list these input values in Table~\ref{tab:04}, as well as the associated outcome of our numerical fit. Note that we use a freedom to freely rotate $10_{Fi}$ and $\overline{5}_{Fj}$ representations in the $SU(5)$ space to go into basis where $Y_1$ of Eqs.\ \eqref{ME} and \eqref{MD} is a diagonal matrix with real parameters. We perform a simultaneous fit of the charged lepton, down-type quark, and neutrino sectors. Whereas, in the up-type quark sector, due to $M_U=M_U^T$, the unitary matrices $U_c$ and $U$ are completely determined via Eq.\ \eqref{eq:U_c}. Of course, what interests us the most are the explicit values of $Y_{11}^{dd}$ and $Y_{11}^{ud}$ as these are crucial input parameters for the strength of the $n-\overline{n}$ oscillation signal.

We investigate the viability of both the normal ordering (NO) and inverted ordering (IO) scenarios for the neutrino masses and associated mixing parameters. Also, we present all our fits in a way that $\mathrm{max}(|Y_{2ij}|)=\mathrm{max}(|Y_{\nu ij}|)=1$, for $i,j=1,2,3$, as the entries in $Y_2$ and $Y_\nu$ can scale freely with $v_{45}$ and $v_{15}$, respectively. Note that we study the $n-\overline{n}$  oscillations only after we perform the fit and check for consistency with the flavor physics constraints of Table \ref{tab:constraints}. All numerical fits are performed with both $P$ and $Q$ of Eq.\ \eqref{eq:U_c} set to be identity matrices, i.e., with $\omega_i=0$, for $i=1,\ldots,5$, to speed up a procedure. We denote our benchmark fit scenarios as Model A (NO) and Model A (IO) depending on the implemented ordering of the neutrino mass parameters. Again, all the input values of the fermion masses and mixing parameters as well as the results of the fit are summarized in Table~\ref{tab:04} for convenience.

\vspace{3pt}\noindent
\textbf{$\bullet$ Model A (NO) benchmark fit:}
\begin{align}
&Y_1=\left(
\begin{array}{ccc}
1.3684752 \times 10^{-6} & 0 & 0 \\
0 & 1.07234 \times 10^{-5} & 0 \\
0 & 0 & 9.14325 \times 10^{-3} \\
\end{array}
\right),
\\
&Y_2=\begin{pmatrix}
0.04368\, e^{-0.42643i} & 0.04211\, e^{0.17418i} & 0.53368\, e^{1.54868i} \\
0.00097\, e^{3.14136i} & 0.10053\, e^{-2.94176i} & 0.02185\, e^{0.02706i} \\
0.24740\, e^{-1.72329i} & 0.66821\, e^{1.05058i} & 1.00000\, e^{0.61401i}
\end{pmatrix},
\\
&Y_\nu=\begin{pmatrix}
0.30619\, e^{-2.97735i} & 0.25071\, e^{-3.13882i} & 0.05531\, e^{-0.02854i} \\
0.25071\, e^{-3.13882i} & 1.00000\, e^{-2.23310i} & 0.80237\, e^{3.12855i} \\
0.05531\, e^{-0.02854i} & 0.80237\, e^{3.12855i} & 0.51987\, e^{3.13948i}
\end{pmatrix},
\\
&v_{45}=0.914813\,\mathrm{GeV},
\quad
v_{15}=0.0162149\,\mathrm{eV}.
\end{align}
Model A (NO) parameter set gives the following elements of $Y^{dd}$ and $Y^{ud}$:
\begin{align}
&Y^{dd}=\begin{pmatrix}
0.5160\, e^{1.7476i} & 0.4641\, e^{1.1119i} & 0.2127\, e^{0.2517i} \\
0.4641\, e^{1.1119i} & 1.5927\, e^{-0.4950i} & 0.9481\, e^{-2.2034i} \\
0.2127\, e^{0.2517i} & 0.9481\, e^{-2.2034i} & 0.6666\, e^{-2.6527i}
\end{pmatrix},
\\
&Y^{ud}=
\begin{pmatrix}
0.0208\, e^{0.9978i} & 0.0706\, e^{0.7180i} & 0.3497\, e^{0.1300i} \\
0.0486\, e^{1.4090i} & 0.1078\, e^{-0.9352i} & 0.2656\, e^{1.0054i} \\
0.1185\, e^{-0.2162i} & 0.8000\, e^{1.8611i} & 0.9697\, e^{0.6558i}
\end{pmatrix}.
\end{align}

The outcome of our numerical fit for the Model A (NO) scenario is specified in a third column of Table \ref{tab:04}. The neutrino sector parameters of interest, namely, the sum of neutrino masses $\sum_i m_i$, the neutrinoless double beta decay parameter $m_{\beta\beta}=| \sum_i (U_\mathrm{PMNS})^2_{ei}m_i |$, and the CP-violating Dirac phase in the leptonic sector $\delta^\mathrm{PMNS}_\mathrm{CP}$, are also summarized in Table \ref{tab:04}. The corresponding experimental bounds/preferences  are also listed in the same Table. 

\vspace{3pt}\noindent
\textbf{$\bullet$ Model A (IO) benchmark fit:}
\begin{align}
&Y_1=\left(
\begin{array}{ccc}
 8.65507 \times 10^{-5} & 0 & 0 \\
 0 & 1.0125 \times 10^{-4} & 0 \\
 0 & 0 & 9.23545 \times 10^{-3} \\
\end{array}
\right),
\\
&Y_2=\begin{pmatrix}
0.04425\, e^{-2.6667i} & 0.00187\, e^{1.1367i} & 0.04697\, e^{0.5744i} \\
0.11784\, e^{2.0829i} & 0.00831\, e^{1.3530i} & 0.33920\, e^{-3.0378i} \\
0.97480\, e^{1.8339i} & 0.30699\, e^{-1.3313i} & 1.00000\, e^{-0.0003i}
\end{pmatrix}
,
\\
&Y_\nu=\begin{pmatrix}
0.00733\, e^{-1.9703i} & 0.89460\, e^{-1.5893i} & 0.26554\, e^{-0.8797i} \\
0.89460\, e^{-1.5893i} & 0.51101\, e^{-2.2655i} & 1.00000\, e^{1.2730i} \\
0.26554\, e^{-0.8797i} & 1.00000\, e^{1.2730i} & 0.50408\, e^{1.2847i}
\end{pmatrix},
\\
&v_{45}=0.769101\,\mathrm{GeV},
\quad
v_{15}=0.0171908\,\mathrm{eV}.
\end{align}
Model A (IO) parameter set gives the following elements of $Y^{dd}$ and $Y^{ud}$:
\begin{align}
&Y^{dd}=\begin{pmatrix}
0.8304\, e^{-2.4130i} & 1.0212\, e^{1.3837i} & 1.3256\, e^{1.8226i} \\
1.0212\, e^{1.3837i} & 0.9656\, e^{2.9352i} & 0.6025\, e^{2.8089i} \\
1.3256\, e^{1.8226i} & 0.6025\, e^{2.8089i} & 0.7516\, e^{0.9952i}
\end{pmatrix},
\\
&Y^{ud}=\begin{pmatrix}
0.0214\, e^{1.7803i} & 0.0446\, e^{-1.4974i} & 0.2710\, e^{-2.8181i} \\
0.0663\, e^{-2.7706i} & 0.0555\, e^{1.9229i} & 0.2918\, e^{-1.7198i} \\
0.7358\, e^{-0.2945i} & 0.9087\, e^{-1.5955i} & 0.8011\, e^{-0.0189i}
\end{pmatrix}.
\end{align}

The outcome of our numerical fit for the Model A (IO) scenario is specified in a fourth column of Table \ref{tab:04}. Note that this scenario  can be fully tested by near future neutrino observables. Namely, neutrino masses are predicted to be 
$\left( m_1,m_2,m_3 \right) =\left( 49.64, 50.39, 5.36 \right) \,\mathrm{meV}$,
which leads to somewhat large value for $\sum_i m_i$, which is close to the ruled out bound from cosmological data~\cite{ParticleDataGroup:2024cfk} that suggests $\sum m_i < 87$\,meV or $\sum m_i < 120$\,meV, depending on experiments included. Model A (IO) fit furthermore yields  $m_{\beta\beta}=48.02$\,meV, which will  also be probed in very near future. 
Furthermore, it is exciting to point out that even though we do not attempt to fit the CP violating phase in the leptonic sector, yet, our benchmark fit corresponds to $\delta^\mathrm{PMNS}_\mathrm{CP}=290^\circ$ which is well within the $1\,\sigma$ range of the current experimental global fit of all neutrino observables. 

\FloatBarrier
\begin{table}[th!]
\centering
\footnotesize
\resizebox{0.95\textwidth}{!}{
\begin{tabular}{|c|c|c|c|}
\hline
\pbox{10cm}{ \textbf{Model A}  } & Experimental values 
(NO/IO)& Fit values (\textbf{NO}) & Fit values (\textbf{IO})   \\ [1ex] \hline\hline

$y_{d}/10^{-5}$ & $6.56\pm0.65$& 6.53787& 6.3507\\ \hline
$y_{s}/10^{-4}$ & $1.24\pm0.06$& 1.24158& 1.25522 \\ \hline
$y_{b}/10^{-2}$ & $0.57\pm0.005$& 0.513947& 0.570674 \\ \hline\hline

$y_{e}/10^{-6}$  &$2.70341\pm0.00270$&2.70331 & 2.7126  \\ \hline
$y_{\mu}/10^{-4}$ &$5.70705\pm0.00570$& 5.71032& 5.7088\\ \hline
$y_{\tau}/10^{-2}$   &$0.97020\pm0.00097$& 0.969925& 0.96683\\ \hline\hline

$\Delta m^2_{21}/10^{-5} \mathrm{eV}^2$ &$7.49\pm 0.19$& 7.47333& 7.4408 \\ \hline
$\Delta m^2_{31/32}/10^{-3} \mathrm{eV}^2$   &$2.5345\pm 0.024$/$-2.484\pm 0.02$& 2.53583& -2.51016\\ \hline\hline

$\sin^2\theta^{\rm{PMNS}}_{12}$  &$0.307^{+0.012}_{-0.011}$& 0.306659& 0.308115\\ \hline
$\sin^2\theta^{\rm{PMNS}}_{23}/10^{-2}$  &$0.561^{+0.012}_{-0.015}$/$0.562^{+0.012}_{-0.015}$& 0.465298& 0.564869\\ \hline
$\sin^2\theta^{\rm{PMNS}}_{13}/10^{-3}$ &$0.02195^{+0.00054}_{-0.00058}$/$0.02224^{+0.00056}_{-0.00057}$& 0.0219057& 0.0225132 \\ 
\hline

\multicolumn{4}{l}{Fit outcomes:} \\   \hline

$\left( m_1,m_2,m_3 \right)$  (meV)   & - &  $\left( 10.27, 13.42, 51.39 \right)$     & $\left( 49.64, 50.39, 5.36 \right)$ \\ \hline

$\sum_i m_i$ (meV)  & <($87$\textendash$120$)~\cite{ParticleDataGroup:2024cfk} &   75.08 &  105.39 \\ \hline

$m_{\beta\beta}$ (meV)  &<($28$\textendash$122$)~\cite{KamLAND-Zen:2024eml} & 10.13  & 48.02  \\ \hline

$\delta^\mathrm{PMNS}_\mathrm{CP}$ (deg) & $177^{+19}_{-20}$/$285^{+25}_{-28}$ \cite{NUFIT} & 74.7& 290.489  \\
[0.5ex]\hline

\end{tabular}
}
\caption{Charged fermion mass parameter input at $M_\mathrm{GUT}$ is  taken from Ref.\  \cite{Babu:2016bmy}. Neutrino observables are taken from Refs.\ \cite{Esteban:2024eli,NUFIT}. 
The Model A (NO) fit yields $\chi^2=1.53$, whereas Model A (IO) fit gives $\chi^2=0.7$.   Some of the interesting fit outcomes relevant for current and future neutrino experiments are also summarized in this Table. These quantities are the sum of neutrino masses $\sum_i m_i$, the neutrinoless double beta decay parameter $m_{\beta\beta}=| \sum_i (U_\mathrm{PMNS})^2_{ei}m_i |$, and the CP-violating Dirac phase in the leptonic sector $\delta^\mathrm{PMNS}_\mathrm{CP}$.}
\label{tab:04}
\end{table}

\begin{figure}[t!]
\centering
\includegraphics[width=0.48\textwidth]{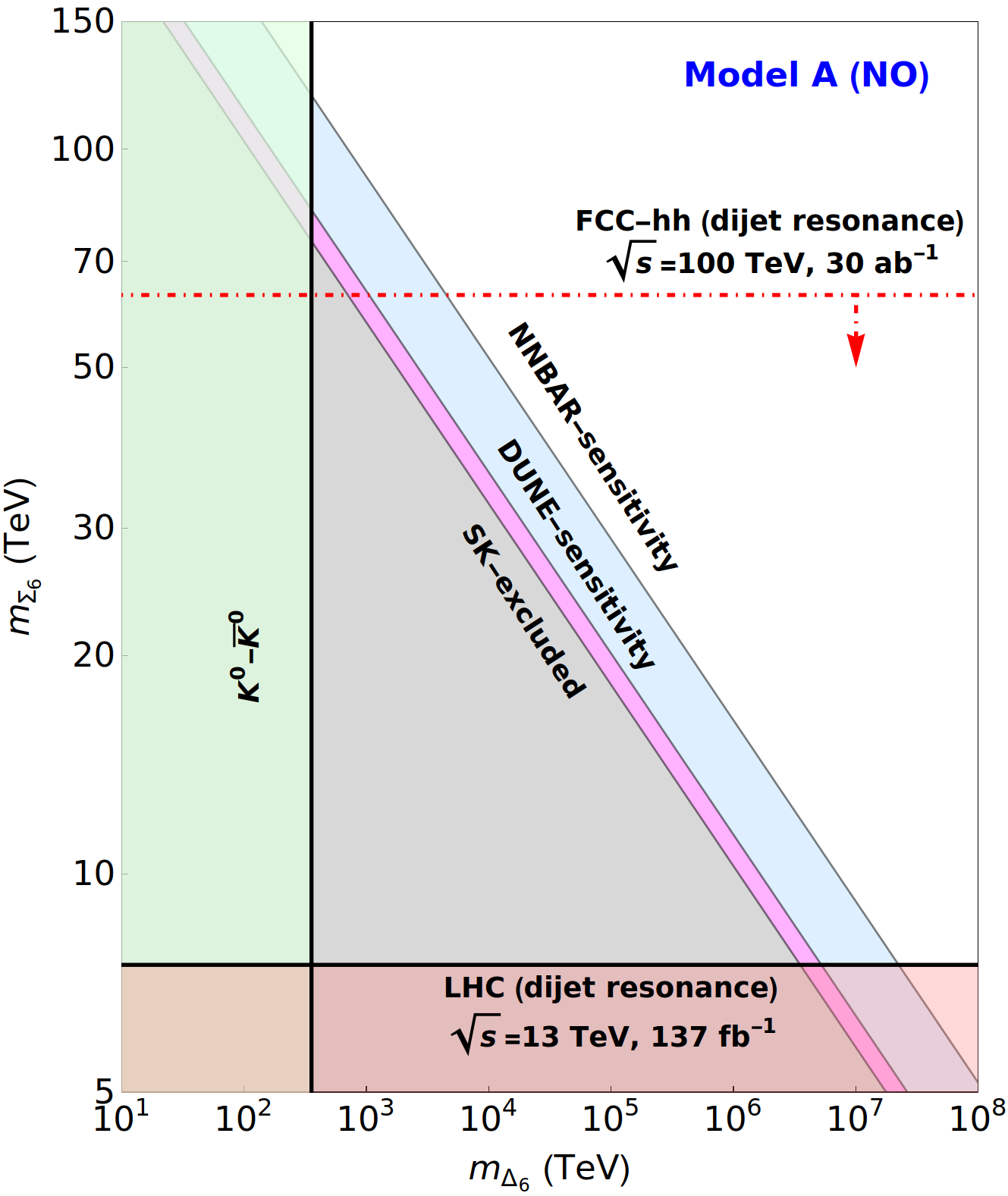}
\includegraphics[width=0.48\textwidth]{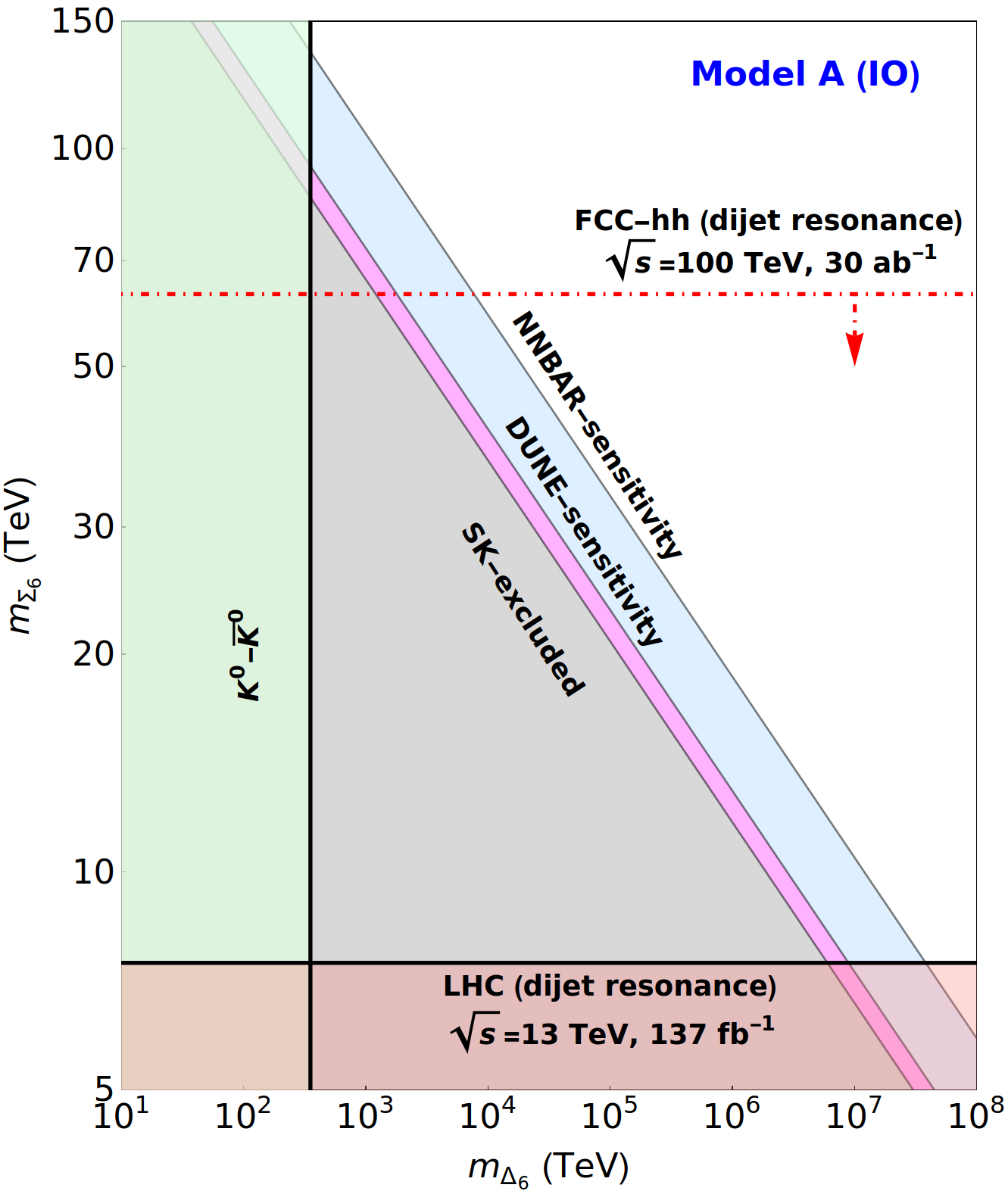}
\caption{ Observable $n-\overline{n}$ range for the Model A (NO) and Model A (IO) scenarios, as indicated. The shaded blue and magenta regions correspond to the viable parts of the parameter space that can be probed in the future experiments, namely, through NNBAR and DUNE experiments, respectively, whereas the shaded gray region plotted on top of the magenta and blue regions shows the current SK exclusion. Light-green vertical shaded region shows the part excluded by $K^0-\overline K^0$ oscillation. Moreover, the pink shaded part highlights the current LHC bound due to dijet resonance.  Projected sensitivity for dijet resonances from FCC-hh is presented by the dash-dotted red line, which will probe a significant part of the existing parameter space.   NO/IO here refers to normal/inverted mass ordering for neutrinos. Both panels are obtained for $\tilde{\mu}=m_{\Delta_6}$. } \label{fig:nnPLOT-A}
\end{figure}

The observable ranges for the $n-\overline{n}$ oscillation signal, as given by Eq.~\eqref{eq:fitnnBAR:A}, for both the Model A (NO) and Model A (IO) scenarios are presented in the left and right panels of Fig.~\ref{fig:nnPLOT-A}, respectively. Again, we set $\tilde{\mu}$ of Eq.\ \eqref{eq:vertex2} to be equal to $m_{\Delta_6}$ to generate Fig.~\ref{fig:nnPLOT-A}. Note that the fact that we require $\mathrm{max}(|Y_{2ij}|)=\mathrm{max}(|Y_{\nu ij}|)=1$, for $i,j=1,2,3$, generates the most conservative parameter space for the observation of $n-\overline{n}$ oscillations. Namely, if one wants to make the relevant entries in either $Y^{uu}$ or $Y^{ud}$ smaller, such a rescaling would shift the blue and magenta regions in both panels of Fig.~\ref{fig:nnPLOT-A} towards lower values of $m_{\Delta_6}$ and $m_{\Sigma_6}$.

We have already mentioned that the phenomenologically viable gauge coupling unification requires $\Sigma_5(3,3,-1/3)\in 45_H$ to be significantly below the unification scale if both $\Sigma_6$ and $\Delta_6$ scalars are light. As the mass ranges of both $\Sigma_6$ and $\Delta_6$ are explicitly given in Fig.~\ref{fig:nnPLOT-A}, we can finally quantify this statement by performing the one-loop level gauge coupling unification analysis with the following SM input parameters: $M_Z=91.1876$\,GeV, $\alpha_S(M_Z)=0.1193\pm0.0016$, $\alpha^{-1}(M_Z)=127.906\pm0.019$, and $\sin^2 \theta_W(M_Z)=0.23126\pm0.00005$~\cite{Agashe:2014kda}.

For example, successful gauge coupling unification in Model A can take place for $(m_{\Delta_6},m_{\Sigma_6})=(350\,\mathrm{TeV},100\,\mathrm{TeV})$ only when $m_{\Sigma_5} \leq 4\times 10^{9}$\,GeV, where all other fields in $15_H$, $24_H$, and $45_H$, except for proton decay mediating scalars, are taken to be free parameters between 1\,TeV and $M_\mathrm{GUT}$. The particular point $(m_{\Delta_6},m_{\Sigma_6})=(350\,\mathrm{TeV},100\,\mathrm{TeV})$ is chosen as it corresponds, for all practical purposes, to the leftmost point of intersection between lines denoted with NNBAR-sensitivity and $K^0-\overline K^0$ in both panels of Fig.~\ref{fig:nnPLOT-A}. (The lower value for the proton decay mediating scalar masses is taken to be $10^{12}$\,GeV.) Note, however, even if one sets $m_{\Sigma_5} = 4\times 10^{9}$\,GeV, the unification scale still comes out to be too low in Model A to be viable as it reads $M^\mathrm{max}_\mathrm{GUT}=1.2 \times 10^{15}$\,GeV, whereas the associated gauge coupling constant is $\alpha_\mathrm{GUT}=1/33.1$. Only if $m_{\Sigma_5}$ is additionally lowered, the unification scale can go sufficiently up for Model A to be viable. Namely, if we again take $(m_{\Delta_6},m_{\Sigma_6})=(350\,\mathrm{TeV},100\,\mathrm{TeV})$, set $m_{\Sigma_5} = 1$\,TeV, and allow all other fields in $15_H$, $24_H$, and $45_H$, except for proton decay mediating fields, to be free parameters between 1\,TeV and $M_\mathrm{GUT}$, we find that the maximal possible value of unification scale is $M^\mathrm{max}_\mathrm{GUT} = 5.1 \times 10^{18}$ GeV while the associated gauge coupling constant is $\alpha_\mathrm{GUT}=1/24.0$. One can say that if $m_{\Sigma_5}$ goes down by two orders of magnitude, $M^\mathrm{max}_\mathrm{GUT}$ goes up by one order of magnitude in Model A.

If, on the other hand, we take a point $(m_{\Delta_6},m_{\Sigma_6})=(3\times 10^7\,\mathrm{TeV},7.5\,\mathrm{TeV})$ that roughly corresponds to the rightmost intersection between lines denoted with NNBAR-sensitivity and LHC (dijet resonance) in both panels of Fig.~\ref{fig:nnPLOT-A}, the unification can take place only if $m_{\Sigma_5} \leq 2\times 10^{11}$\,GeV. For $m_{\Sigma_5} = 2\times 10^{11}$\,GeV we find $M_\mathrm{GUT}^\mathrm{max} = 1.0 \times 10^{15}$ GeV while $\alpha_\mathrm{GUT}=1/34.1$. If we set $m_{\Sigma_5} = 1$\,TeV, we find $M^\mathrm{max}_\mathrm{GUT}=8.5 \times 10^{18}$\,GeV, whereas $\alpha_\mathrm{GUT}=1/25.0$. Again, all other scalars in $15_H$, $24_H$, and $45_H$, in this analysis, are allowed to go as low as 1\,TeV to maximise GUT except for the proton decay mediating fields that are kept at or above $10^{12}$\,GeV.

Finally, we note that although the flavor physics puts a constraint on $m_{\Sigma_6}$, especially data on $\Delta m_{B_d}$, this constraint on $m_{\Sigma_6}$, for the $Y^{ud}$ entries used, is weaker than the LHC limit of 7.5\,TeV. It is thus not displayed in Fig.~\ref{fig:nnPLOT-A}. 

\section{Model B}\label{sec:03}

\subsection{Neutrino mass}

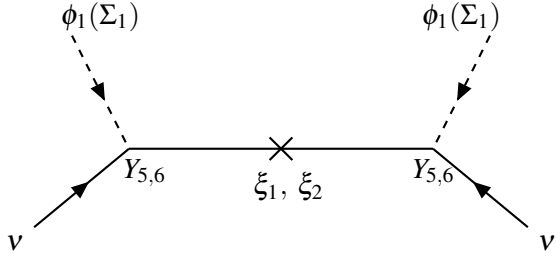
\begin{figure}[t]
\centering
\begin{tikzpicture}
\begin{feynman}
\vertex (a3) at (-3.5,-1.25) {\(\nu \)};
\vertex (b3) at (3.5,-1.25) {\(\nu \)};
\coordinate (v1) at (-2.75,1.5);
\coordinate (v2) at (-2,0);
\coordinate (v3) at (2.75,1.5);
\coordinate (v4) at (2,0);
\coordinate (mid) at (0,0);

\diagram* {
  (v1) -- [charged scalar, thick] (v2),
  (a3) -- [fermion, thick] (v2),
  (v3) -- [charged scalar, thick] (v4),
  (b3) -- [fermion, thick] (v4),
  (v2) -- [plain, thick] (v4),
};

\node at ($(v1)!0.5!(v2) + (0,1)$) {\(\phi_1 (\Sigma_1)\)};
\node at ($(v3)!0.5!(v4) + (0,1)$) {\(\phi_1 (\Sigma_1)\)};
\draw[thick] (mid) ++(-0.15,0.15) -- ++(0.3,-0.3);
\draw[thick] (mid) ++(-0.15,-0.15) -- ++(0.3,0.3);

\node at (0.1,-0.5) {\(\xi_1,\;\xi_2\)};
\end{feynman}
\node at (-2.2,0) [below right] {\( Y_{5,6} \)};
\node at (1.6,0) [below right] {\( Y_{5,6} \)};
\end{tikzpicture} \caption{Neutrino mass through type I ($\xi_1(1,1,0)$) and type III  ($\xi_2(1,3,0)$)  contributions.}\label{fig:nuMASS-B}
\end{figure}

To generate neutrino masses within Model B, one requires a single 24-dimensional fermionic representation $24_F$. Its decomposition reads
\begin{align}
24_F&=\xi_1(1,1,0)+\xi_2(1,3,0)+\xi_3(8,1,0) +
   \xi_4(3,2,-5/6)+\overline\xi_4(\overline{3},2,5/6),
\end{align} 
where the relevant fields for the neutrino mass generation are $\xi_1 (1,1,0)$ and $\xi_2 (1,3,0)$. The contractions of interest for our discussion on neutrino mass generation are
\begin{align}
\mathcal{L}_Y\supset Y_{5 i}\; \overline 5_{F i} 24_F 5_H +   Y_{6 i}\;  \overline 5_{F i} 24_F 45_H + m_{24}\; 24^2_F  +\lambda \;  24^2_F 24_H, 
\label{eq:lagrangian_B}
\end{align}
where we omit $SU(5)$ indices for simplicity.
The associated neutrino mass diagram is shown in Fig.\ \ref{fig:nuMASS-B}.

The last two terms in Eq.\ \eqref{eq:lagrangian_B} contribute to the mass of the SM multiplets comprising $24_F$. Therefore, there are two mass relations that read 
\begin{align}
&m_{\xi_1}=\frac{3}{5} m_{\xi_2}   +\frac{2}{5} m_{\xi_3}, 
\quad
m_{\xi_4}=\frac{1}{2} m_{\xi_2}   +\frac{1}{2} m_{\xi_3}.
\label{eq:24Fmassrelations}
\end{align}

The first two terms of Eq.\ \eqref{eq:lagrangian_B}, on the other hand, provide couplings to neutrinos, which are
\begin{align}
\mathcal{L}_\nu &\supset  \nu_i \xi^0_1 v_5\; \underbrace{ \bigg\{
\frac{1}{2}\sqrt{\frac{3}{10}}  Y_{5 i} - \frac{\sqrt{5}}{8} \frac{v_{45}}{v_5} Y_{6 i}
\bigg\}   }_{\widetilde Y_i} 
+  \nu_i \xi^0_2 v_5\; \underbrace{ \bigg\{
\frac{1}{2\sqrt{2}}  Y_{5 i} + \frac{\sqrt{3}}{8} \frac{v_{45}}{v_5} Y_{6 i}
\bigg\}   }_{\hat Y_i} .
\end{align}
Since $\widetilde Y$ and $\hat Y$ are three-vectors, the neutrino mass matrix $M_N$ takes the following form:
\begin{align}
M_N= \frac{v_5^2}{m_{\xi_1}}\ \widetilde Y^T \widetilde Y  +    \frac{v_5^2}{m_{\xi_2}} \ \hat Y^T \hat Y. 
\end{align}

For potentially observable $n-\overline n$ oscillations $\xi_3$ needs to be relatively light, as we demonstrate later on. Therefore, we can make the following approximation 
\begin{align}
&m_{\xi_1}=\frac{3}{5} m_{\xi_2}   +\frac{2}{5} m_{\xi_3} \sim \frac{3}{5} m_{\xi_2}.
\end{align}
The neutrino mass matrix can thus be written as  
\begin{align}
M_N= \frac{5v_5^2}{3m_{\xi_2}}\ \widetilde Y^T \widetilde Y  +    \frac{v_5^2}{m_{\xi_2}} \ \hat Y^T \hat Y. 
\end{align}

\subsection{ $n-\overline{n}$\, Oscillation}

The $n-\overline{n}$ oscillation process proceeds through a diagram shown in Fig.\ \ref{fig:nnBAR-B} that corresponds to Topology B of Fig.\ \ref{fig:topology}. The origin of the $n-\overline{n}$ transition in Model B lies in the existence of a scalar color sextet and a color-octet fermion in the theory. The scalar sextet is required to provide the necessary corrections to the charged fermion masses, while the color-octet fermion arises from implementation of the Type III seesaw mechanism responsible for viable neutrino mass generation. It is important to note that the realization of this diagram requires an adjoint fermion, which in our case is the $(8,1,0)$ component contained within the adjoint fermion $24_F$ of $SU(5)$.

The interaction strength of the two lower vertices of Fig.\ \ref{fig:nnBAR-B} is generated by the interaction 
\begin{align}
&\mathcal{L}_Y\supset Y_{6 i}\; \overline 5_{F i} 24_F 45_H    \supset \overline Y_i\; d^c_i \xi_3 \Sigma_6 \;\varepsilon_c,
\end{align}
where $\overline Y_i= \frac{1}{2} \left( D^T_c Y_6^T\right)_i$ is written in the physical basis ($\varepsilon_c$ is the three-index Levi-Civita tensor). The gauge invariance of the term on the right-hand side can be understood from the fact that, under the color $SU(3)_c$ group, $\overline{6} \times \overline{3} \supset 8$.   Of course, it is only the value of $\overline Y_1$ that matters for our purposes. The upper two vertices in Fig.\ \ref{fig:nnBAR-B}, on the other hand, are defined in Eq.\ \eqref{eq:Yud}.
\begin{figure}[t]
\centering
\begin{tikzpicture}
\begin{feynman}

\vertex (a1) at (-0.8,2.5) {\(d^c \)};
\vertex (a2) at (-3.2,2.5) {\(u^c \)};
\vertex (b1) at (0.9,2.5) {\(u^c \)};
\vertex (b2) at (3.2,2.5) {\(d^c \)};
\vertex (a3) at (-3,-1) {\(d^c \)};
\vertex (b3) at (3,-1) {\(d^c \)};

\coordinate (v1) at (-2,1.5);
\coordinate (v2) at (-1.5,0);
\coordinate (v3) at (2,1.5);
\coordinate (v4) at (1.5,0);
\diagram* {
  (a1) -- [fermion, thick] (v1), 
  (v2) -- [charged scalar, thick, edge label={\(\Sigma^*_6\)}] (v1),
  (a2) -- [fermion, thick] (v1),
  (a3) -- [fermion, thick] (v2),
  (b1) -- [fermion, thick] (v3),
  (v4) -- [charged scalar, thick, edge label'={\(\Sigma^*_6\)}] (v3),
  (b2) -- [fermion, thick] (v3),
  (b3) -- [fermion, thick] (v4),
  (v2) -- [plain, thick, edge label] (v4),
};
\node at (0.1,-0.5) {\(\xi_3 \)};
\node at (-1.5,-0.4) {\( \overline Y_1 \)};
\node at (1.5,-0.4) {\( \overline Y_1 \)};
\node at (2,2.2) {\(  Y^{ud}_{11} \)};
\node at (-2,2.2) {\(  Y^{ud}_{11} \)};
\end{feynman}
\end{tikzpicture} \caption{ $n-\overline{n}$ oscillation diagram in Model B.  }\label{fig:nnBAR-B}
\end{figure}

The strength of the
$\Delta B=2$ operator resulting from Fig.\ \ref{fig:nnBAR-B} can be estimated to be
\begin{align}
\mathcal{L}_\mathrm{eff} \sim \frac{(Y^{ud}_{11})^2(\overline Y_{1})^2}{m_{\xi_3}m_{\Sigma_6}^4}.    
\end{align}
Therefore, $\tau^{-1}_{n-\overline{n}}$ reads
\begin{align}
    \tau^{-1}_{n-\overline{n}}= \langle \overline n |\mathcal{L}^{n-\overline{n}}_\mathrm{eff}| n \rangle  = \left| \frac{(Y^{ud}_{11})^2(\overline Y_{1})^2}{m_{\xi_3}m_{\Sigma_6}^4}  \mathcal{M}(2\,\mathrm{GeV})\right|,
\label{eq:modelBtau}
\end{align}
where the relevant transition matrix element typically has the value $\mathcal{M} \sim 10^{-5}\,\mathrm{GeV}^6$. (See, for example, Table 1 of Ref.\ \cite{Fridell:2021gag} for more details.) If we assume $m_{\xi_3} = m_{\Sigma_6} = M$ and set all relevant Yukawa couplings to one, Eq.\ \eqref{eq:modelBtau} yields $\tau_{n-\overline n}\gtrsim 4.7\times 10^8\,\mathrm{sec}$ for $M \gtrsim 3.7\times 10^5\,\mathrm{GeV}$. Finally, if we demand an observability of the $n-\overline{n}$ oscillation signal in near future, the relevant range for Model B reads
\begin{align}
&
4.56\times 10^{13} \; \mathrm{(NNBAR)} \geq 
\frac{1}{ |Y^{ud}_{11}|^2\left|\overline Y_{1}\right|^2 } \left( \frac{m_{\xi_3}}{ \mathrm{1\,TeV} }  \right)  \left( \frac{ m_{\Sigma_6}}{ \mathrm{1\,TeV}} \right)^4  
   \geq    7.14\times 10^{12}\; \mathrm{(SK)}. \label{eq:fitnnBAR:B}  
\end{align}

What we address next is whether Model B is compatible with the window of observability presented in Eq.\ \eqref{eq:fitnnBAR:B}.

\subsection{Numerical analysis}

\subsubsection{Flavor constraints}
We note that both the mass and the couplings of $\Sigma_6$ scalar can be constrained by the one-loop level meson-antimeson oscillations presented in Fig.\ \ref{mmBAR-loop}. Therefore, we impose conditions for $\Delta m_{B_s}$, $\Delta m_{B_d}$, and $\Delta m_{K}$ from Table~\ref{tab:constraints} that place constraints on $\sum_{i=1}^3 \left|Y^{ud}_{i3}{Y^{ud}_{i2}}^*\right|$, $\sum_{i=1}^3 \left|Y^{ud}_{i3}{Y^{ud}_{i1}}^*\right|$, and $\sum_{i=1}^3 \left|Y^{ud}_{i1}{Y^{ud}_{i2}}^*\right|$, respectively. We find, for the benchmark fits provided, that these flavor constraints are automatically satisfied.

\subsubsection{Numerical fit}
We follow the same procedure that was used to analyze Model A to provide the benchmark fit parameters for Model B. We, again, consider both the NO and IO scenarios for the neutrino mass and mixing parameters that we denote with Model B (NO) and Model B (IO), respectively. Also, all the input values of the fermion masses and mixing parameters as well as the results of the fits are summarized in Table~\ref{tab:05} for convenience.
All numerical fits are performed with both $P$ and $Q$ of Eq.\ \eqref{eq:U_c} set to be identity matrices, while the solutions that we present are such that $\mathrm{max}(|Y_{2ij}|)=\mathrm{max}(|Y_{6i}|)=1$, for $i,j=1,2,3$.

\vspace{3pt}\noindent
\textbf{$\bullet$ Model B (NO) benchmark fit:}
\begin{align}
&Y_1=\left(
\begin{array}{ccc}
1.17497 \times 10^{-4} & 0 & 0 \\
0 & 2.38081 \times 10^{-4} & 0 \\
0 & 0 & 8.94557 \times 10^{-3} \\
\end{array}
\right),
\\
&Y_2=\begin{pmatrix}
0.01664\, e^{-3.12734i} & 0.03585\, e^{-2.33885i} & 1.00000\, e^{-0.09239i} \\
0.00378\, e^{2.60508i} & 0.19694\, e^{-0.38734i} & 0.98382\, e^{2.39164i} \\
0.03138\, e^{0.01263i} & 0.15024\, e^{-2.67314i} & 0.83700\, e^{-0.11357i}
\end{pmatrix},
\\
&Y_5^T=\begin{pmatrix}
0.0014775 \, e^{-2.7607 i} \\
0.0077954 \, e^{1.2267 i} \\
0.0069110 \, e^{-0.8066 i}
\end{pmatrix}, \;\; Y_6^T=\begin{pmatrix}
0.88515\, e^{-1.50703i} \\
0.81606\, e^{-0.11981i} \\
1.00000\, e^{0.64138i}
\end{pmatrix},
\\
&v_{45}=0.834365\,\mathrm{GeV},
\quad m_{\xi_2}= 3.23585\times 10^{10}\,\mathrm{GeV}.
\end{align} 
This parameter set gives the following entries that are relevant for the $n-\overline n$ oscillation signal:
\begin{align}
&\overline Y=\begin{pmatrix}
0.1862\, e^{-2.9696 i} \\
0.5683\, e^{0.7806 i} \\
0.5046\, e^{0.6208 i}
\end{pmatrix}, 
\;
&Y^{ud}=\begin{pmatrix}
0.0568\, e^{-1.317 i} & 0.0750\, e^{-1.7228 i} & 1.1626\, e^{-0.2194 i} \\
0.1425\, e^{1.3692 i} & 0.1570\, e^{0.7592 i} & 0.8104\, e^{2.2397 i} \\
0.1316\, e^{-0.8576 i} & 0.1040\, e^{-1.5099 i} & 0.8010\, e^{-0.0959 i}
\end{pmatrix}.
\end{align}

Our benchmark fit corresponds to $\delta^\mathrm{PMNS}_\mathrm{CP}=190^\circ$, which is, interestingly, well within the $1\,\sigma$ range of the experimental global fit of all neutrino observables.

\vspace{5pt}\noindent
\textbf{$\bullet$ Model B (IO) benchmark fit:}
\begin{align}
&Y_1=\left(
\begin{array}{ccc}
 1.20531\times 10^{-4} & 0 & 0 \\
 0 & 2.56473\times 10^{-4} & 0 \\
 0 & 0 & 8.95859\times 10^{-3} \\
\end{array}
\right),
\\
&Y_2=\begin{pmatrix}
0.0168058\, e^{-3.12773 i} & 0.0365247\, e^{-2.38752 i} & 0.979713\, e^{-0.09030 i} \\
0.00387949\, e^{2.59567 i} & 0.193765\, e^{-0.40711 i} & 1.00000\, e^{2.38227 i} \\
0.0322388\, e^{0.01224 i} & 0.150992\, e^{-2.67040 i} & 0.814236\, e^{-0.11662 i}
\end{pmatrix},
\\
&Y_5^T=\begin{pmatrix}
0.00405208\, e^{-1.95762 i} \\
0.00115246\, e^{-2.84840 i} \\
0.00223453\, e^{-2.40660 i}
\end{pmatrix}, \;\; Y_6^T=\begin{pmatrix}
0.9103\, e^{-0.6751 i} \\
0.9758\, e^{2.4789 i} \\
1.0000\, e^{2.4148 i}
\end{pmatrix},
\\
&v_{45}=0.843558\mathrm{GeV},
\quad m_{\xi_2}=4.6329 \times 10^{14} \mathrm{GeV}.
\end{align}
Derived quantities from the fit relevant for computing $n-\overline{n}$ oscillation are:
\begin{align}
&\overline Y=\begin{pmatrix}
0.6063\, e^{-1.6612 i} \\
0.2516\, e^{2.7411 i} \\
0.5141\, e^{2.3963 i}
\end{pmatrix}\, ,
\quad Y^{ud}=\begin{pmatrix}
0.0593\, e^{-1.2892 i} & 0.0741\, e^{-1.7109 i} & 1.1446\, e^{-0.2229 i} \\
0.1455\, e^{1.3863 i} & 0.1509\, e^{0.7595 i} & 0.8302\, e^{2.2344 i} \\
0.1359\, e^{-0.8257 i} & 0.1004\, e^{-1.4859 i} & 0.7778\, e^{-0.0974 i}
\end{pmatrix}.
\end{align}

\begin{table}[th!]
\centering
\footnotesize
\resizebox{0.95\textwidth}{!}{
\begin{tabular}{|c|c|c|c|}
\hline
\pbox{10cm}{ \textbf{Model B}  } & Experimental values (NO/IO) & Fit values  (\textbf{NO})  & Fit values  (\textbf{IO}) \\ [1ex] \hline\hline

$y_{d}/10^{-5}$ & $6.56\pm0.65$& 6.72908&6.63154 \\ \hline
$y_{s}/10^{-4}$ & $1.24\pm0.06$& 1.22096&1.24446  \\ \hline
$y_{b}/10^{-2}$ & $0.57\pm0.005$& 0.567719&0.570248  \\ \hline\hline

$y_{e}/10^{-6}$  &$2.70341\pm0.00270$&2.69724&2.70262  \\ \hline
$y_{\mu}/10^{-4}$ &$5.70705\pm0.00570$& 5.7190& 5.70852\\ \hline
$y_{\tau}/10^{-2}$   &$0.97020\pm0.00097$& 0.970562&0.969377 \\ \hline\hline

$\Delta m^2_{21}/10^{-5} \mathrm{eV}^2$ &$7.49\pm 0.19$& 7.48031&7.46139  \\ \hline
$\Delta m^2_{31/32}/10^{-3} \mathrm{eV}^2$   &$2.5345\pm 0.024$/$-2.484\pm 0.02$& 2.53579&-2.50572 \\ \hline\hline

$\sin^2\theta^{\rm{PMNS}}_{12}$  &$0.307^{+0.012}_{-0.011}$& 0.307845&0.308784 \\ \hline
$\sin^2\theta^{\rm{PMNS}}_{23}/10^{-2}$  &$0.561^{+0.012}_{-0.015}$/$0.562^{+0.012}_{-0.015}$& 0.561693& 0.565515\\ \hline
$\sin^2\theta^{\rm{PMNS}}_{13}/10^{-3}$ &$0.02195^{+0.00054}_{-0.00058}$/$0.02224^{+0.00056}_{-0.00057}$& 0.0219064& 0.0223054 \\ \hline\hline

\multicolumn{4}{l}{Fit outcomes:} \\   \hline

$\left( m_1,m_2,m_3 \right)$  (meV)   & - &  $\left( 0, 8.648, 50.35 \right)$     & $\left( 49.306, 50.05, 0 \right)$ \\ \hline

$\sum_i m_i$ (meV)  & <($87$\textendash$120$)~\cite{ParticleDataGroup:2024cfk} &   59 &  59 \\ \hline

$m_{\beta\beta}$ (meV)  &<($28$\textendash$122$)~\cite{KamLAND-Zen:2024eml} & 3.69  & 48.4  \\ \hline

$\delta^\mathrm{PMNS}_\mathrm{CP}$ (deg) & $177^{+19}_{-20}$/$285^{+25}_{-28}$ \cite{NUFIT} & 191.67 & 174.039 \\
[0.5ex]\hline

\end{tabular}
}
\caption{ Charged fermion mass parameter input at $M_\mathrm{GUT}$ is taken from Ref.\  \cite{Babu:2016bmy}. Neutrino observables are taken from Refs.\ \cite{Esteban:2024eli,NUFIT}. The Model B (NO) fit yields $\chi^2=1.0$, whereas Model B (IO) fit gives $\chi^2=0.13$. }
\label{tab:05}
\end{table}

\begin{figure}[t!]
\centering
\includegraphics[width=0.48\textwidth]{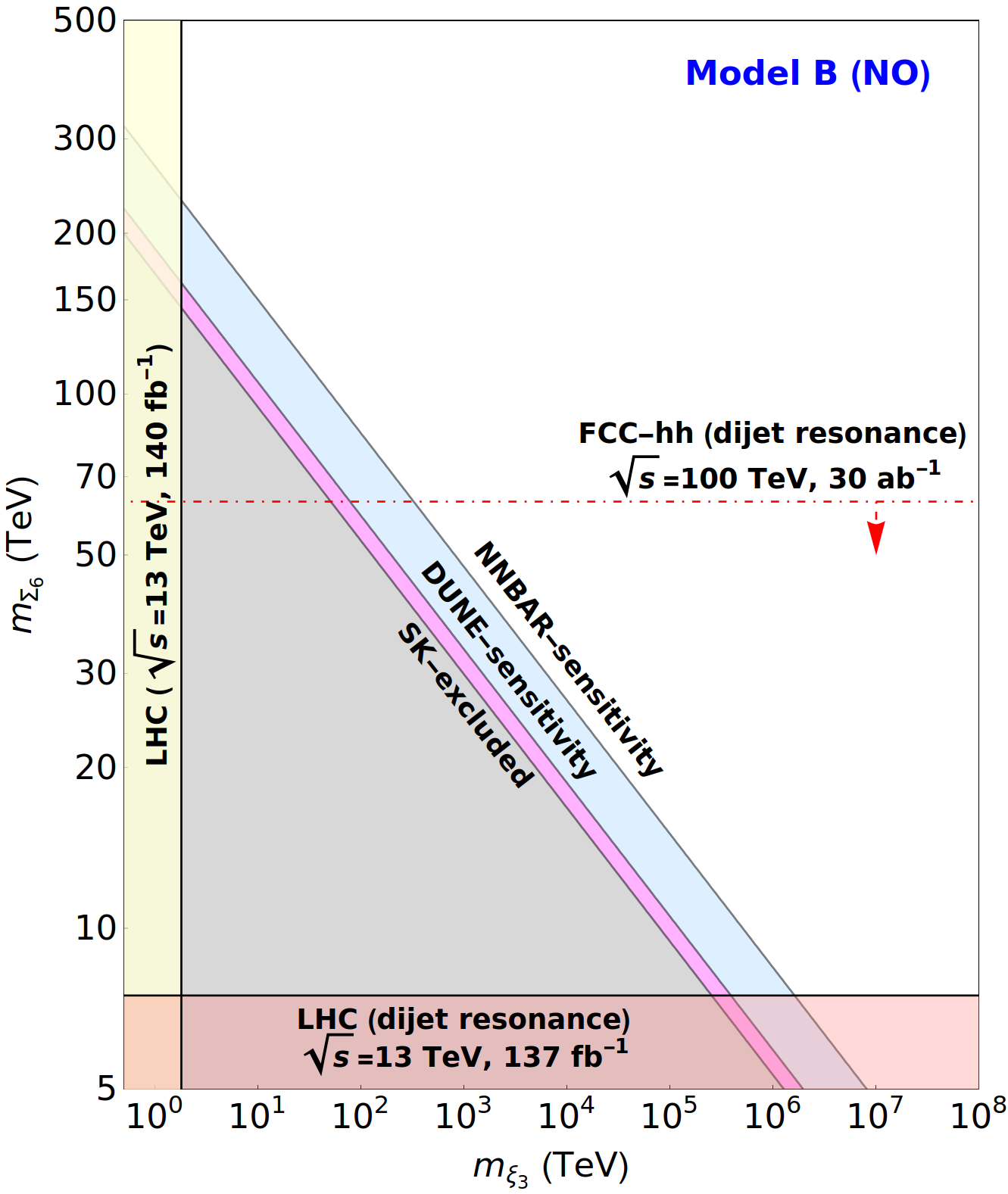}
\includegraphics[width=0.48\textwidth]{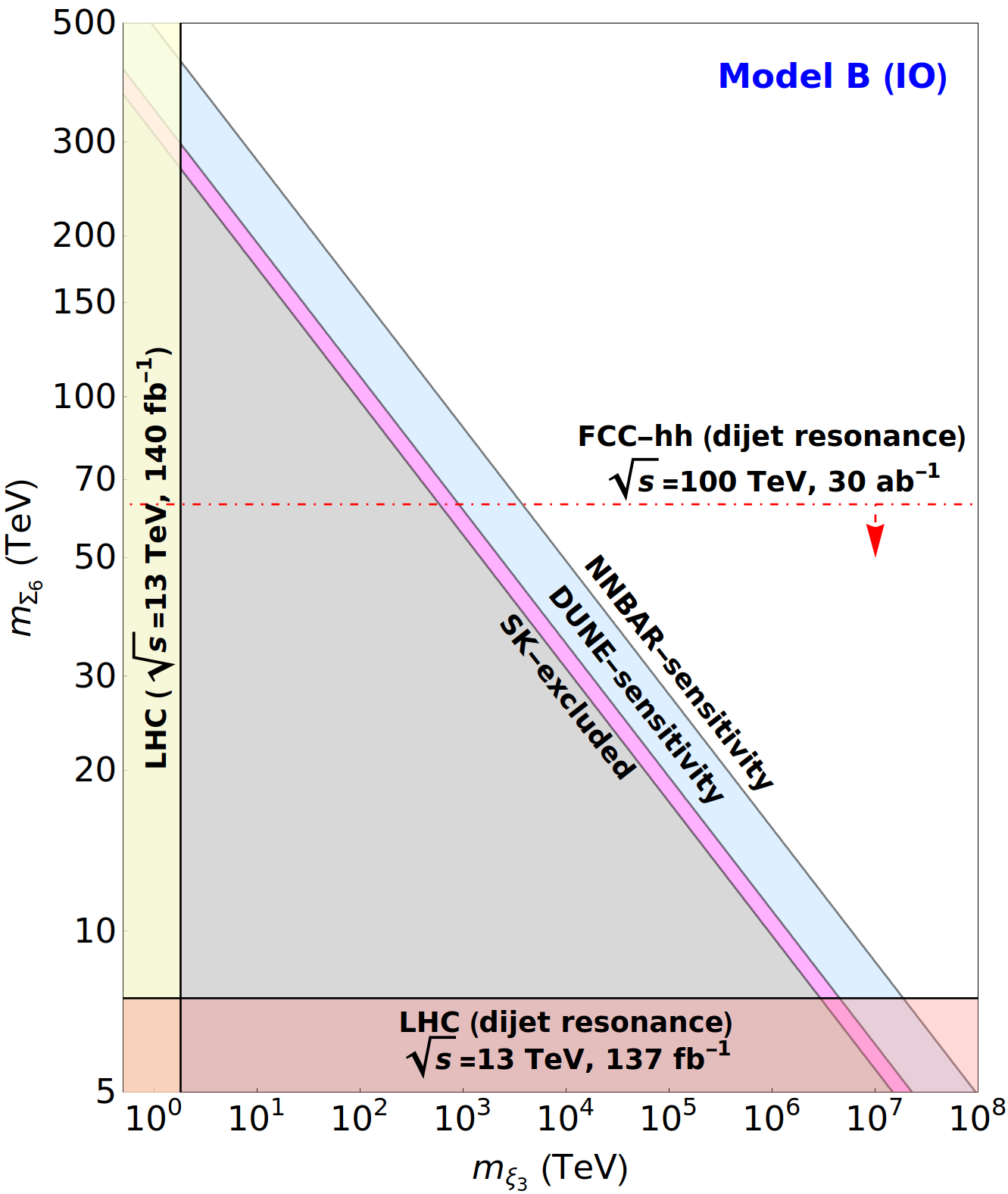}
\caption{   Observable  $n-\overline{n}$ ranges for the Model B (NO) and Model B (IO) scenarios.  The shaded blue and magenta regions correspond to the viable parts of the parameter space that can be probed in the future by NNBAR experiment and DUNE, respectively, whereas the shaded gray region plotted on top of magenta and blue regions shows the current exclusion from SK. The pink shaded part highlights the current LHC bound due to dijet resonance, whereas the vertical yellow shaded region corresponds to exclusion from pair-produced octet colored fermions, each decaying into three jets~\cite{ATLAS:2024kqk}.  Projected sensitivity for dijet resonances from FCC-hh is presented by the dash-dotted red line, which will probe a significant part of the existing parameter space.  } \label{fig:nnPLOT-B}
\end{figure}

Predictions for both benchmark scenarios within Model B framework are presented in Fig.\ \ref{fig:nnPLOT-B}.
The pink shaded part in Fig.~\ref{fig:nnPLOT-B} highlights the current LHC bound due to dijet resonance. Vertical yellow shaded region is a constraint due to pair-produced octet colored fermions, where each fermion decays into three jets, thus mimicking R-parity violating gluino signals~\cite{ATLAS:2024kqk}.

To address gauge coupling unification within Model B, we take two particular points in the $m_{\xi_3}$-$m_{\Sigma_6}$ plane of Fig.~\ref{fig:nnPLOT-B}. The first one is $(m_{\xi_3},m_{\Sigma_6})=(1.8\,\mathrm{TeV},300\,\mathrm{TeV})$ that corresponds, for all practical purposes, to the leftmost point of intersection between line denoted with NNBAR-sensitivity and the yellow exclusion region boundary in both panels of Fig.~\ref{fig:nnPLOT-B}. The second one is $(m_{\xi_3},m_{\Sigma_6})=(10^7\,\mathrm{TeV},7.5\,\mathrm{TeV})$ and it corresponds to the rightmost intersection between lines denoted with NNBAR-sensitivity and LHC (dijet resonance) in both panels of Fig.~\ref{fig:nnPLOT-B}.

Successful gauge coupling unification in Model B, for $(m_{\xi_3},m_{\Sigma_6})=(1.8\,\mathrm{TeV},300\,\mathrm{TeV})$, can only take place when $m_{\Sigma_5} \leq 3\times 10^{5}$\,GeV, where all other fields in $24_F$, $24_H$, and $45_H$, except for proton decay mediating scalars, are taken to be free parameters between 1\,TeV and $M_\mathrm{GUT}$. For the fields in $24_F$ we also implement mass relation of Eq.\ \eqref{eq:24Fmassrelations}. If one sets $m_{\Sigma_5} = 3\times 10^{5}$\,GeV, the unification scale is $M^\mathrm{max}_\mathrm{GUT}=1.7 \times 10^{16}$\,GeV, whereas the associated gauge coupling is $\alpha_\mathrm{GUT}=1/25.3$. If $m_{\Sigma_5}$ is lowered, the unification scale goes up. For example, if we again take $(m_{\xi_3},m_{\Sigma_6})=(1.8\,\mathrm{TeV},300\,\mathrm{TeV})$, set $m_{\Sigma_5} = 1$\,TeV, and allow all other fields in $24_F$, $24_H$, and $45_H$, except for proton decay mediating fields, to be free parameters between 1\,TeV and $M_\mathrm{GUT}$, we find that the maximal possible value of unification scale is $M^\mathrm{max}_\mathrm{GUT} = 3.7 \times 10^{18}$\,GeV while the associated gauge coupling constant is $\alpha_\mathrm{GUT}=1/14.0$. One can say that if $m_{\Sigma_5}$ goes down by one order of magnitude, $M^\mathrm{max}_\mathrm{GUT}$ goes up by one order of magnitude in Model B.

For point $(m_{\xi_3},m_{\Sigma_6})=(10^7\,\mathrm{TeV},7.5\,\mathrm{TeV})$ the unification can take place only if $m_{\Sigma_5} \leq 10^{5}$\,GeV. For example, if $m_{\Sigma_5} = 10^{5}$\,GeV, we find $M_\mathrm{GUT}^\mathrm{max} = 2.5 \times 10^{16}$ GeV while $\alpha_\mathrm{GUT}=1/32.8$. If, on the other hand, we set $m_{\Sigma_5} = 1$\,TeV, we find $M^\mathrm{max}_\mathrm{GUT}=3.3 \times 10^{18}$\,GeV, whereas $\alpha_\mathrm{GUT}=1/22.5$. Again, all other fields in $24_F$, $24_H$, and $45_H$, in this analysis, are allowed to go as low as 1\,TeV to maximise GUT except for the proton decay mediating fields that are kept at or above $10^{12}$\,GeV.

This discussion on gauge coupling unification in Model B confirms that $\Sigma_5(3,3,-1/3)\in 45_H$ has to be significantly below the unification scale if both $\Sigma_6$ and $\xi_3$ are light, in agreement with our initial claim.

Before concluding, we summarize the common features and the distinctive aspects of the two models discussed in this paper:

\begin{itemize}
\item In both models realistic charged fermion masses are achieved through the presence of a $45_H$, which contains a crucial $ud$-type color sextet $\Sigma_6$. This field plays an essential role in generating $n-\overline{n}$ oscillations.

\item Model A employs the type II seesaw mechanism for neutrino mass generation, realized by  $\Delta_1(1,3,1)$ scalar field residing within the $15_H$ representation. Interestingly, the same representation also contains a $dd$-type scalar sextet $\Delta_6$, which, together with the $ud$-type sextet from the $45_H$, facilitates $n-\overline{n}$ oscillations.

\item Model B, in contrast, makes use of the type III seesaw mechanism via the $24_F$ fermion multiplet, which inherently includes a color-octet $(8,1,0)$ fermion $\xi_3$. This new fermionic state, in conjunction with the $ud$-type sextet from the $45_H$, gives rise to $n-\overline{n}$ transitions.

\item While both models generate $n-\overline{n}$ transitions, the underlying interaction structures differ. In Model A, it is a cubic scalar interaction involving three color sextets that contributes towards $n-\overline n$ oscillation, whereas in Model B the relevant contributions arise solely from Yukawa interactions.

\item In both models, the color sextets are constrained by current LHC searches to lie  approximately above $7.5$\,TeV. However, in Model A the additional $dd$-type sextet introduces severe flavor-violating effects, particularly meson–antimeson oscillations, which push the constraints to the scale of  $\mathcal{O}(100)$\,TeV. By contrast, in Model B the absence of such sextets implies that the lower limits on beyond the SM states leading to $n-\overline{n}$ oscillations are dominated entirely by collider bounds.

\item Both models allow for either normal or inverted mass ordering of neutrinos. However, Model B uniquely predicts the lightest neutrino to be massless.

\item These scenarios, specified by the benchmark fits presented above, can be tested in upcoming neutrino experiments through their predictions for neutrino mixing parameters, notably at DUNE~\cite{DUNE:2020jqi}, JUNO~\cite{JUNO:2015zny}, and Hyper-Kamiokande~\cite{Hyper-Kamiokande:2018ofw}. In addition, future searches for neutrinoless double-beta decay, such as nEXO~\cite{nEXO:2021ujk}, will fully probe the relevant parameter space in both models. Moreover,  the sum of neutrino masses is already stringently constrained by cosmological measurements~\cite{ParticleDataGroup:2024cfk}, and, in the near future, the inverted mass ordering scenario for Model B may be fully ruled out.

\item  For the models presented in this work, future searches for $n-\overline{n}$ oscillations at DUNE and at the NNBAR facility at ESS are expected to achieve unprecedented sensitivities, probing new-physics scales up to $10^{10}$ GeV and $10^{11}$ GeV, respectively, while the other state involved in the process needs to be near the TeV scale.   

\item   It should be understood that the non-observation of $n-\overline{n}$ oscillation would not rule out the model. It would only indicate that the states responsible for this $\Delta B = 2$ process are relatively heavy, leading to a suppression of the $n-\overline{n}$ transition rate.  
\end{itemize}

\section{Conclusions}\label{sec:04}
Neutron-antineutron oscillations offer a powerful probe of baryon number violation and high-scale new physics well beyond the Standard Model. Future experiments such as DUNE and the proposed NNBAR at ESS are poised to extend sensitivity to such $\Delta B = 2$ processes at unprecedented levels, probing energy scales far beyond the reach of colliders. Motivated by this, we have presented two simple $SU(5)$-based models in which the embedding of the seesaw mechanism for neutrino masses naturally realizes two distinguishable topologies for $n-\overline{n}$ oscillations. One is mediated by color-sextet scalars (arising in a Type II seesaw context), and the other one involves both a scalar sextet and a color-octet fermion (emerging from a Type III seesaw structure). While the first topology can also arise within the $SO(10)$/Pati–Salam framework, the second embeds naturally in $SU(5)$, this distinction constitutes a key novelty of our proposal. In both models, the same dynamics responsible for generating fermion masses also induces baryon number violation, linking flavor structure and baryon number violation in a unified framework.   Moreover, we have shown that if $n-\overline{n}$ oscillations are observed, the FCC-hh’s projected sensitivity to dijet resonances will probe a substantial portion of the allowed parameter space.  These findings highlight the unique potential of $n-\overline{n}$ oscillation searches to access ultra-heavy new physics and offer a rare low-energy portal to grand unification. We also demonstrated that a consistent gauge coupling unification, in agreement with current proton decay bounds,   is achieved within these models in the parameter space where $n-\overline{n}$ oscillations are potentially observable. Gauge-mediated proton decay, if observed, would offer a complementary probe for testing these models.

\subsection*{Acknowledgments}
I.D.\ acknowledges the financial support from the Slovenian Research Agency (research core funding No.\ P1-0035). S.F.\ and  S.S.\ acknowledge the financial support
from the Slovenian Research Agency (research core funding No.\ P1-0035 and N1-0321).

\bibliographystyle{style}
\bibliography{reference}
\end{document}